\documentclass[aps,prl,reprint,superscriptaddress]{revtex4-1}
\usepackage{natbib}
\usepackage{amsmath}
\usepackage[toc,page]{appendix}
\usepackage{bm}
\usepackage{ amssymb }
\usepackage{graphicx}
\usepackage {epstopdf}
\usepackage{csquotes}

\begin{document}


\title{Direct observation of anyonic braiding statistics at the $\nu$=1/3 fractional quantum Hall state}



\author{J. Nakamura}
\affiliation{Department of Physics and Astronomy, Purdue University}
\affiliation{Birck Nanotechnology Center, Purdue University}


\author{S. Liang}
\affiliation{Department of Physics and Astronomy, Purdue University}
\affiliation{Birck Nanotechnology Center, Purdue University}

\author{G. C. Gardner}
\affiliation{Birck Nanotechnology Center, Purdue University}
\affiliation{Microsoft Quantum Purdue, Purdue University}

\author{M. J. Manfra}
\email[]{mmanfra@purdue.edu}
\affiliation{Department of Physics and Astronomy, Purdue University}
\affiliation{Birck Nanotechnology Center, Purdue University}
\affiliation{School of Electrical and Computer Engineering, Purdue University}
\affiliation{Microsoft Quantum Purdue, Purdue University}
\affiliation{School of Materials Engineering, Purdue University}

\date{\today}

\begin{abstract}
Utilizing an electronic Fabry-Perot interferometer in which Coulomb charging effects are suppressed, we report experimental observation of anyonic braiding statistics for the $\nu=1/3$ fractional quantum Hall state. Strong Aharonov-Bohm interference of the $\nu=1/3$ edge mode is punctuated by discrete phase slips consistent with an anyonic phase of $\theta_{anyon}=\frac{2\pi}{3}$. Our results are consistent with a recent theory of a Fabry-Perot interferometer operated in a regime in which device charging energy is small compared to the energy of formation of charged quasiparticles \cite{Rosenow2019}. Close correspondence between device operation and theoretical predictions substantiates our claim of observation of anyonic braiding.
\end{abstract}

\maketitle

\section{Background}
Quantum theory requires that all fundamental particles must be fermions or bosons, which has profound implications for particles' statistical behavior. However, theoretical works have shown that in two dimensions it is possible for particles to violate this principle and obey so-called anyonic statistics, in which exchange of particle position results in a quantum mechanical phase change that is not $\pi$ or $2\pi$ (as for fermions or bosons), but a rational fraction of $\pi$  \cite{Leinaas1977, Wilczek1982}. While anyons cannot exist as fundamental particles in nature, certain condensed matter systems are predicted to host exotic quasiparticles which obey a certain form of anyonic statistics. 

The quantum Hall effect is a remarkable example of a topological phase of matter occurring when a two-dimensional electron system (2DES) is cooled to low temperature and placed in a strong magnetic field. In the quantum Hall regime the bulk forms an insulator, and charge flows in edge currents which are topologically protected from backscattering and exhibit quantized conductance. The elementary excitations of fractional quantum Hall states \cite{Tsui1982} are not simply electrons, which obey fermionic statistics, but instead are emergent quasiparticles which are predicted to have highly exotic properties including fractional charge and anyonic statistics \cite{Laughlin1983}. In two dimensions, two exchanges of particle positions are topologically equivalent to one quasiparticle encircling the other in a closed path \cite{Nayak2008}, referred to as a braid; this is illustrated in Fig. \ref{Device}a. The anyonic character of these quasiparticles is reflected in the fractional phase the system obtains from braiding; thus they are said to obey anyonic braiding statistics. The statistics of fractional quantum Hall states have been studied in theoretical \cite{Halperin1984, Arovas1984} and numerical \cite{Kjonsberg1999, Kjonsberg1999-2, Jain2003, Jain2004, JainBook} works. The anyonic phase does not depend on the trajectory taken but only on the number of quasiparticles encircled, making braiding another manifestation of topology in quantum Hall physics; this topological robustness has motivated aggressive pursuit of fault-tolerant quantum computation based on braiding operations in various condensed matter systems \cite{DasSarma2005, Nayak2008, Stern2013, DasSarma2015}. In a recent experimental work anyonic statistics were inferred from noise correlation measurements \cite{Bartolomei2020}; however, direct observation of the anyonic phase in braiding experiments will further our understanding of the exotic behavior of quantum Hall quasiparticles and is a necessary step to towards quasiparticle manipulation.

Electronic interferometry has been used to study edge physics in previous theoretical  \cite{Rosenow2019, Chamon1997, Halperin2007, Halperin2011, Rosenow2012, Levkivskyi2012, VonKeyserlingk2015, Gefen2016, Frigeri2019} and experimental \cite{Heiblum2003, Roulleau2008, Litvin2008, Deviatov2008, Zhang2009, McClure2009, Goldman2009, Deviatov2011, Heiblum2010, Huynh2012, Ensslin2013, Heiblum2015, Heiblum2016, Heiblum2016-2, Tewari2016, Heiblum2017, Heiblum2018, Duprez2019, Heiblum2019, Ensslin2020} works, and has been proposed as an experimental means to observe anyonic braiding statistics \cite{Kivelson1990, Chamon1997, Kim2006, Halperin2011} including the highly exotic non-Abelian form of anyonic statistics \cite{Halperin2006, Bonderson2006, Bishara2008, Bishara2009, Willett2009, Stern2010, Willett2013, Willett2019}. An electronic Fabry-Perot interferometer consists of a confined 2DES using quantum point contacts (QPCs) to partition edge currents, as shown in Fig. \ref{Device}b. Quasiparticles backscattered by the QPCs will braid around quasiparticles localized inside the interferometer; therefore changes in $N_{qp}$, the number of quasiparticles localized inside the interferometer, will result in a shift in the interference phase due to the anyonic contribution $\theta_{anyon}$ \cite{Kivelson1990, Chamon1997, Kim2006, Halperin2011}, with $\theta_{anyon}=\frac{2\pi}{2p+1}$ for a Laughlin fractional quantum Hall state $\nu=\frac{1}{2p+1}$ \cite{Halperin1984, Arovas1984}. The interferometer phase difference $\theta$ is a combination of the Aharonov-Bohm phase scaled by the quasiparticle charge $e^*$ and the anyonic contribution,  written in Eqn. \ref{FractAB} \cite{Kivelson1990, Chamon1997, Halperin2011}:

\begin{equation} \label{FractAB}
\theta = 2\pi \frac{e^*}{e} \frac{A_I  B}{\Phi _0}+N_{qp} \theta _{anyon}
\end{equation}

The total current backscatterd by the interferometer will depend on  $\cos{(\theta)}$, so the interference phase can be probed by measuring the conductance $G$ across the device \cite{Nakamura2019}. 

A major obstacle towards the observation of anyonic phases through interferometry has been been the Coulomb interaction of the interfering edge state with charge located in the bulk of the interferometer \cite{Halperin2007}. A strong bulk-edge interaction causes the area $A_I$ of the interferometer to change when charge in the bulk changes \cite{Halperin2007, Halperin2011}. As a consequence, for so-called Coulomb-dominated devices with strong bulk-edge interaction, the change in Aharonov-Bohm phase due to the change in $A_I$ when $N_{qp}$ is changed cancels out the anyonic phase $\theta_{anyon}$, making quasiparticle braiding statistics unobservable \cite{Halperin2011}. While novel physics has been explored in Coulomb-dominated devices \cite{Zhang2009, Heiblum2010, Goldman2009, McClure2012, Kou2012, Ensslin2020}, this bulk-edge interaction must be reduced to make anyonic braiding observable. Various techniques have been implemented to reduce this Coulomb bulk-edge interaction, including the use of metal screening gates \cite{Zhang2009, Heiblum2010}, low-temperature illumination to enhance screening by the doping layer \cite{Willett2009, Willett2013, Willett2019}, addition of an Ohmic contact inside the interferometer \cite{Heiblum2016-2}, and incorporation  of auxilliary screening layers inside the semiconductor heterostructure \cite{Nakamura2019}. The screening layer technique has enabled the use of small highly coherent interferometers that exhibit robust Aharonov-Bohm interference, including at fractional quantum Hall states \cite{Nakamura2019}.



\section{Device Design}
\begin{figure}[t]
\def\ffile{MainTextFigures/Device9.pdf}
\centering
\includegraphics[width=\linewidth]{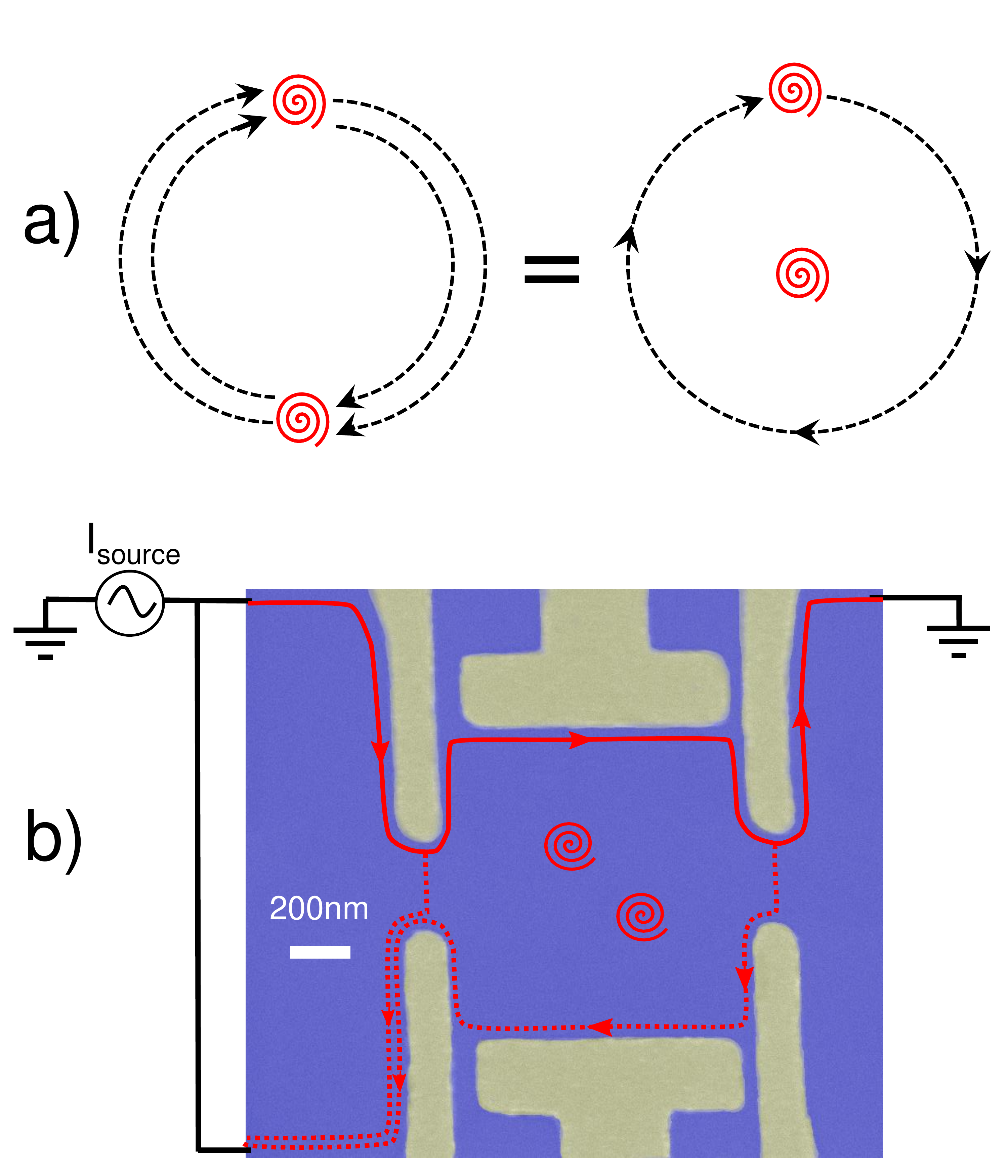}
\caption{\label{Device} \textbf{Quasiparticle braiding experiment.} a) Schematic representation of quasiparticle exchange; quasiparticles are represented by red vortices, and trajectories are shown in dashed lines. Two quasiparticle exchanges (left) which bring the particles back to their original position are topologically equivalent to one quasiparticle executing a closed loop around the the other, and in each case the system gains a quantum mechanical phase $\theta_{anyon}$ due to the quasiparticle's anyonic braiding statistics. b) False-color SEM image of interferometer. Blue regions indicate the GaAs where the 2DES resides, and metal gates under which the 2DES is depleted are highlighted in yellow. When operated at the $\nu = 1/3$ fractional quantum Hall state, the current is carried by quasiparticles traveling in chiral edge states (red arrows), and dotted arrows indicate the backscattered quasiparticle paths which may interfere. Quasiparticles may be localized inside the chamber of the interferometer, as represented by the red vortices, and the backscattered paths enclose a loop around these quasiparticles, making the interferometer sensitive to $\theta_{anyon}$. The lithographic area is 1.0$\mu$m $\times$ 1.0$\mu$m. The device used in the experiments also has a metal gate covering the top of the interferometer not shown in b), which is kept at ground potential and does not affect the 2DES density underneath.}
\end{figure}

The device used for these experiments utilizes a unique high-mobility GaAs/AlGaAs heterostructure \cite{Manfra2014, Gardner2016} with screening layers to minimize the bulk-edge interaction (see the layer stack in Supp. Fig. 1) \cite{Nakamura2019}. The interferometer is defined using metal surface gates which are negatively biased to deplete the 2DES underneath. Two narrow constrictions define QPCs to backscatter edge currents, and wider side gates define the rest of the interference path. An SEM image of the device is shown in Fig. \ref{Device}b; the device has a nominal area of 1.0$\mu$m $\times$ 1.0 $\mu$m, and measurements suggest that lateral depletion of the 2DES makes the interferometer area smaller by approximately 200nm on each side, similar to the experimental and numerical results in \cite{Nakamura2019} (see also \cite{Harshad2018}). Note that the length scale of the interferometer is much greater than the magnetic length $l_B\equiv\sqrt{\frac{hc}{eB}}$ in the regime investigated, with $l_B\approx$9nm at $\nu = 1/3$, so the condition that the interfering quasiparticles be well separated from the localized  quasiparticles inside the interferometer which they may braid around should hold \cite{Jain2003, Jain2004}.  Compared to the device used in \cite{Nakamura2019}, the device used in this work has a lower electron density $n$, which improves device stability because smaller gate voltages can be used. The device also has a somewhat smaller area, which may increase coherence and visibility of interference. Experiments are performed in a dilution refrigerator with a base mixing chamber temperature of $T\approx$10mK; Coulomb blockade measurements of different quantum dot devices suggest a somewhat higher electron temperature of $T\approx 22$mK. Negative voltages of $\approx$-1V are applied to the QPC gates and $\approx$-0.8V on the side gates; conductance is measured as a function of the side gate voltage variation $\delta V_g$, which is relative to -0.8V and applied to both side gates. An additional metal gate in the center of the device (not shown in Fig. \ref{Device}b for clarity) is held at ground potential, so it does not affect the 2DES density; this gate is intended to make the confining potential from the gates sharper. Measurements are performed using standard 4-terminal and 2-terminal lock-in amplifier techniques. 

\section{Discrete Phase Slips}

\begin{figure*}[t]
\def\ffile{MainTextFigures/DiscreteJumpsBatch2Scan2v6.pdf}
\centering
\includegraphics[width=\linewidth]{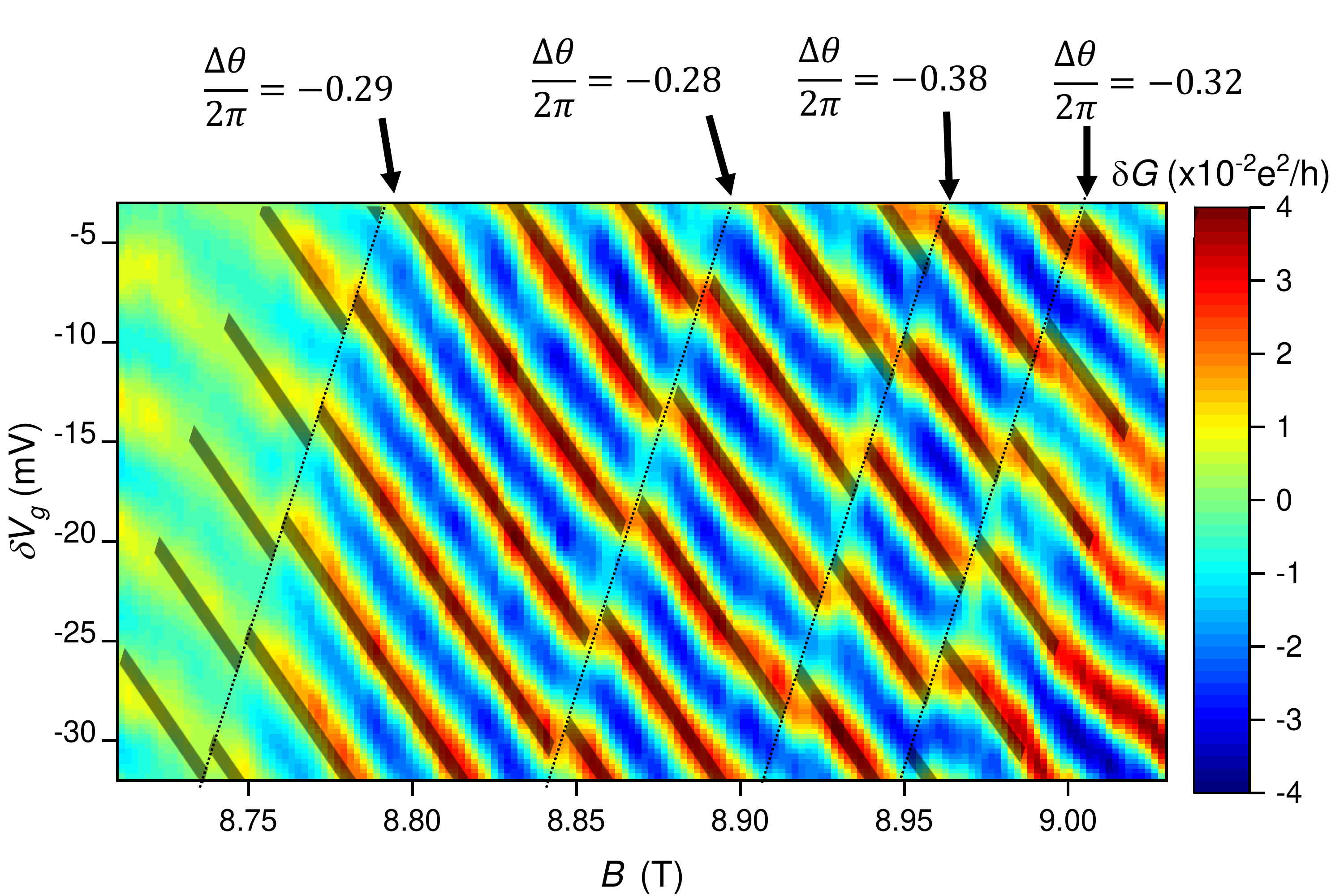}
\caption{\label{DiscreteJumps} \textbf{Conductance oscillations versus magnetic field and side gate voltage}. The predominant behavior is  negatively sloped Aharonov-Bohm interference, but a small number of discrete phase jumps are visible. Dashed lines are guides to the eye for these features. Least-squares fits of $\delta G=\delta G_0 \cos{(2\pi \frac{AB}{\Phi_0}+\theta_0)}$ are shown with highlighted stripes, and the extracted change in phase $\frac{\Delta \theta}{2 \pi}$ are indicated for each discrete jump. Increasing magnetic field is expected to reduce the number of localized quasiparticles; therefore the change in phase across each jump is predicted to be $-\theta _{anyon}$. }
\end{figure*}


We operated the device at high magnetic field $B$ at the filling factor $\nu = 1/3$ quantum Hall state. In Fig. \ref{DiscreteJumps} we show the conductance variation $\delta G$ measured across the interferometer versus $B$ and $\delta V_{g}$ near the center of the $\nu = 1/3$ conductance plateau. The QPCs remain in the regime of weak backscattering across this region with approximately 90$\%$ transmission, and a smooth background conductance is subtracted so that the interference oscillations can be seen clearly. As can be seen in the figure, the predominant behavior observed is conductance oscillations with negatively-sloped lines of constant phase; however, quite conspicuously there are also a small number of discrete phase jumps in the data; dotted lines are guides to the eye for these features. The jumps in phase were found to be repeatable in subsequent scans; see Supp. Fig. 5. 

Eqn. \ref{FractAB} provides a straightforward explanation for our observations. Continuous phase evolution with negatively-sloped lines of constant phase is the signature of the Aharonov-Bohm effect \cite{Halperin2011}. This can be seen by taking the derivative of Eqn. \ref{FractAB} with $\theta$ and $N_{qp}$ held fixed, which yields $\frac{\partial V_g}{\partial B}=-\frac{1}{\beta} \frac{A_I}{B}$ (here $\beta \equiv \frac{\partial A_I}{\partial V_g}$ parameterizes the change in interferometer area with side gate voltage). The negative sign implies negative slope to lines of constant phase as has been observed in previous experiments in the integer \cite{Heiblum2003, Heiblum2010, Zhang2009} and fractional \cite{Nakamura2019} quantum Hall regimes. The second term in Eqn. \ref{FractAB} predicts a discrete change in phase when the number of localized quasiparticles changes; therefore, it is natural to associate the discrete phase jumps with the anyonic phase contribution $\theta_{anyon}$. It is noteworthy that the discrete jumps in phase occur across lines with positive slope in the $B$-$V_g$ plane. This can be understood from the fact that increasing $B$ is expected to remove quasiparticles from the bulk (or create quasiholes) \cite{Laughlin1983, Chamon1997}, while increasing gate voltage would make it electrostatically favorable to increase the number of localized quasiparticles. Thus, the magnetic field at which it becomes favorable to remove a quasiparticle should increase when gate voltage is increased, and a positive slope to the quasiparticle transitions is expected, as observed in resonant tunneling experiments \cite{Chklovskii1996, Goldman2009, McClure2012, Kou2012, Heiblum2017}. The fact that we do indeed observe a positive slope strongly suggests that these discrete phase jumps are associated with changes in localized quasiparticle number, and the magnitude of the slope is also consistent with this; see Supp. Discussion 1 for additional analysis. Furthermore, a central principle of quantum Hall theory is that quasiparticles are localized in the hills and valleys of the disorder potential \cite{Halperin1982}, and the fact that the discrete phase jumps are irregularly spaced indicates that their positions are in fact determined by disorder as expected.

To determine the value of the change in phase associated with each phase jump in the data, we performed a least-squares fit in the regions between the phase jumps, fitting the conductance data to the form $\delta G = \delta G_0 cos(2\pi \frac{1}{3}\frac{A_I B}{\Phi_0}+\theta_0)$, with the fitting parameter being $\theta_0$. This expression for the conductance assumes that between the discrete phase jumps, the phase evolves only by the change in Aharonov-Bohm phase with changing $B$ and changing $A_I$ (via the change in $V_g$), and $\theta _0$ is the excess phase which cannot be attributed to the Aharonov-Bohm effect. We determine the value of the phase jump by computing $\Delta \theta$, the difference in the fitted values of $\theta_0$ in adjacent regions. The fitted data are shown highlighted in Fig. \ref{DiscreteJumps}, and the extracted values of $\frac{\Delta \theta}{2\pi}$ are shown above each jump. Taking an average and assuming that each phase jump corresponds to the removal of a quasiparticle (or equivalently addition of a quasihole), we obtain $\theta _{anyon}={2\pi}\times(0.31 \pm 0.04)$; this is consistent with the theoretical value of $\theta_{anyon}=\frac{2\pi}{3}$ for the $\nu = 1/3$ state \cite{Halperin1984, Arovas1984}. Our work thus provides experimental confirmation for the prediction of fractional braiding statistics at the $\nu = 1/3$ quantum Hall state.


\section{Transition from constant filling to constant density}

\begin{figure*}[t]
\def\ffile{MainTextFigures/PlateauPajamasv6.pdf}
\centering
\includegraphics[width=\linewidth]{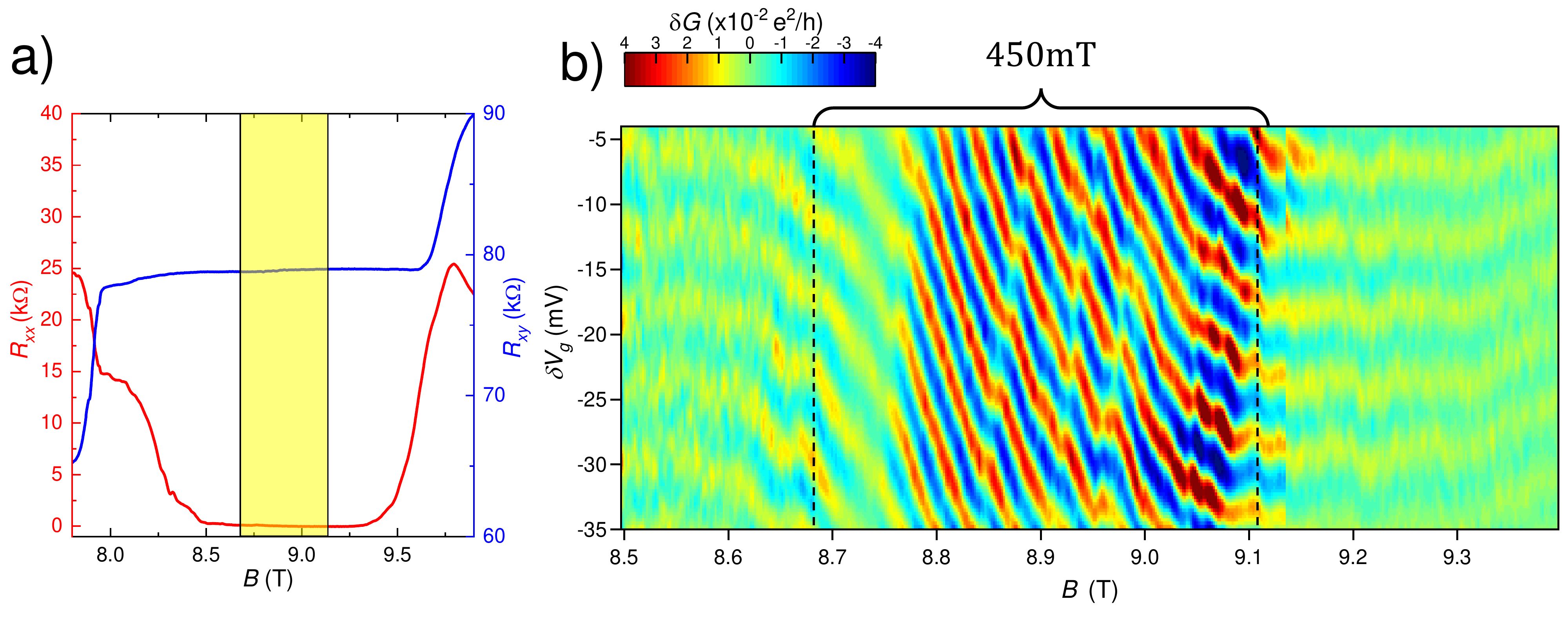}
\caption{\label{PlateauPajamas} \textbf{Interference across the $\nu = 1/3$ quantum Hall  plateau.} a) Bulk magnetransport showing longitudinal resistance $R_{xx}$ and Hall resistance $R_{xy}$ across the $\nu = 1/3$ state. b) Conductance oscillations $\delta G$ versus magnetic field $B$ and side gate voltage $\delta V_{g}$ (this side gate voltage variation is relative to -0.8V). The dashed lines indicate the approximate range over which the device appears to exhibit conventional Aharonov-Bohm interference with minimal influence of the anyonic phase contribution. The region over which this occurs is near the center of the plateau, and is highlighted in the bulk transport data in a).}
\end{figure*}

A recent theoretical work analyzed the case of a Fabry-Perot interferometer operated at the $\nu = 1/3$ state in which strong screening is utilized to reduce the characteristic Coulomb charging energy and thus suppress the bulk-edge interaction \cite{Rosenow2019}. A key prediction is that the device will transition from a regime of constant filling factor to regimes of constant electron density when the magnetic field is varied away from the center of the state and the chemical potential moves away from the center of the gap in the density of states. The authors find that over a wide range of magnetic field the bulk 2DES stays at fixed $\nu = 1/3$ filling. In this regime of constant $\nu$ the predominant contributor to the phase will be the Aharonov-Bohm phase, but a small number of well-separated quasiparticle transitions should occur from which $\theta_{anyon}$ may be extracted, consistent with our results described above. Once the magnetic field is varied away from the center, the authors predict that the electrostatic energy cost of varying density to maintain fixed $\nu$ will cause a transition from constant filling factor to constant density. In the regimes of constant density, many quasiparticles (at low field) or quasiholes (at high magnetic field) will be created inside the interferometer to keep the total charge fixed, with one quasiparticle or quasihole created when the flux is changed by one flux quantum $\Phi_0$, resulting in significant changes in interference behavior mediated by the anyonic phase. 

Motivated by these predictions, we operated the interferometer in a wide range of magnetic field across the $\nu = 1/3$ fractional quantum Hall state. Bulk magnetotransport at $\nu =1/3$ with vanishing longitudinal resistance $R_{xx}$ and a quantized plateau in the Hall resistance $R_{xy}$ is shown in Fig. \ref{PlateauPajamas}a, showing the range of magnetic field over which the $\nu = 1/3$ state occurs in our sample. The conductance measured across the device across the $\nu = 1/3$ state is shown in Fig. \ref{PlateauPajamas}b; this is the same measurement as shown in Fig. \ref{DiscreteJumps}, but extended to higher and lower magnetic field. As  discussed previously, near the center of the $\nu=1/3$ plateau the predominantly observed behavior in the conductance is lines of constant phase with negative slope consistent with Aharonov-Bohm interference \cite{Chamon1997, Halperin2007, Halperin2011, Nakamura2019} with a small number of discrete jumps attributed to quasiparticle transitions. The gate voltage and magnetic field oscillation periods are approximately three times larger than the integer periods measured at $\nu = 1$, consistent with interference of e/3 fractionally charged quasiparticles, as is expected for the $\nu = 1/3$ state and consistent with previous experimental observations of  fractional charge \cite{Goldman1995Science, Heiblum1997, Saminadayar1997, Heiblum2010, Nakamura2019}. On either side of this central region, however, the behavior changes significantly. The lines of constant phase lose their negative slope; although there is still weak magnetic field dependence to the pattern, the magnetic field scale over which the phase varies is much larger than in the central region, making the lines of constant phase nearly flat; the oscillations depend primarily only on the side gate voltage. It is noteworthy that, despite this conspicuous change, the lines of constant phase are continuous across the transition from the central Aharonov-Bohm region to the upper and lower regions, which indicates that the oscillations are still due to interference of the edge state.

Our experimental observation that negatively-sloped Aharonov-Bohm interference occurs only in a finite range of magnetic field agrees with the predictions of \cite{Rosenow2019}. At first blush the behavior observed above and below this central region seems to conflict with predictions: we observe an interference pattern that becomes nearly independent of magnetic field, while \cite{Rosenow2019} predicts that the magnetic field period will decrease from $3\Phi_0$ in the central region to $\Phi_0$ in the upper and lower regions because quasiparticles will be created with period $\Phi_0$. However, an additional key prediction in \cite{Rosenow2019} is that the $\Phi_0$ oscillations will be extremely susceptible to thermal smearing, with the authors estimating a temperature scale $T_0\approx 2$mK (because our device is smaller than the one considered in \cite{Rosenow2019} this predicted temperature scale would be $T_0\approx4$mK for our device, still much smaller than our estimated electron temperature of $T\approx22$mK). This thermal smearing can be understood from the fact that the regime of constant density corresponds to the chemical potential being at a position of high density of states, and thus small energy spacing between states, leading to thermal smearing. Therefore, the absence of $\Phi_0$ oscillations at $T\approx22$mK is in fact in agreement with \cite{Rosenow2019}. 

The fact that the lines of constant phase flatten out and become independent of magnetic field can be understood based on the combined contribution of the Aharonov-Bohm phase and anyonic phase (Eqn. \ref{FractAB}). For the $\nu = 1/3$ state, quasiparticles are predicted to carry fractional charge $e^* = e/3$ and fractional braiding statistics $\theta_{anyon} = 2 \pi /3$ \cite{Arovas1984}. Changing the magnetic field to add one flux quantum to the device will change the Aharonov-Bohm phase by $\frac{2\pi}{3}$. Additionally, in the lower field regimes one quasiparticle will be removed, and in the high field regime one quasihole will be added, resulting in a phase shift of $-\frac{2\pi}{3}$ and leaving the total interference phase unchanged in both regimes. The Aharonov-Bohm phase varies continuously, while (in the limit of zero temperature) the quasiparticle number will change discretely, leading to the predicted $\Phi_0$ oscillations \cite{Chamon1997, Rosenow2019}; however, when the quasiparticle number is thermally smeared, the average number of localized quasiparticles will vary nearly continuously, leading to a smooth variation of the anyonic phase; in this case the smoothly varying thermally-averaged anyonic phase cancels the Aharonov-Bohm phase, leading to no change in $\theta$ as $B$ is varied, consistent with our experimental observations. Because each quasiparticle at the 1/3 state is a vortex, this can also be understood based on the result from \cite{Arovas1984} that the Berry phase of a vortex encircling a closed path is equal to $2 \pi \langle q_{enc} \rangle$ where $q_{enc}$ is the charge enclosed in the path, and the high and low field regions the electrostatics force density to remain fixed, and thus $\langle q_{enc} \rangle$ remains nearly constant.  

The approximate range over which the negatively-sloped Aharonov-Bohm oscillations occur is marked with dashed lines in Fig. \ref{PlateauPajamas}a, and has a span of approximately 450mT. To make a quantitative comparison to theory, we compute the predicted width of the fixed $\nu$ region from \cite{Rosenow2019}: $\Delta B_{constant-  \nu} = \frac{\Delta_{1/3}\Phi _0 C_{SW}  }{\nu e^2 e^*}$. In this expression $\Delta_{1/3}$ is the energy gap of the $\nu = 1/3$ state which we measure to be $\approx 5.5$K (see Supp. Fig. 3), consistent with previous measurements of the $\nu=1/3$ gap \cite{Stormer1993}. $C_{SW}$ is the capacitance per unit area of the screening layers to the quantum well which we calculate as $C_{SW} = \frac{2\epsilon}{d}$, with the factor of two accounting for the fact that there are two screening layers and $d=48$nm the setback of the screening layers from the quantum well. Using the experimental values from the device gives a predicted value for $\Delta B_{fixed- \nu} \approx 530$mT, in good agreement with the experimentally observed range of Aharonov-Bohm interference of $\approx 450$mT, which suggests that the experimentally observed transition in interference behavior can indeed be explained by the model of \cite{Rosenow2019}.  

Additionally, there is a moderate reduction in the side gate voltage oscillation period in the high and low field regions compared to the central region. In Supp. Discussion 1 and Supp. Fig 4 we analyze this shift in period and extract the parameters $\alpha_{bulk}$ relating the change in bulk charge and $\alpha_{edge}$ relating the change in charge at the edge to side gate voltage. Using these parameters, we find that the shift in period is consistent with creation of quasiparticles and quasiholes with gate voltage, leading to a change in period through the statistical phase. Also, we have performed numerical simulations of interferometer behavior based on the models of \cite{Halperin2011} and \cite{Rosenow2019} which show good qualitative agreement with experiment; see Supp. Discussion 2 and Supp. Fig. 5. 

Taken together, our observations of discrete  phase jumps near the center of $\nu = 1/3$ along with the transitions in behavior at high and low field in concordance with the predictions of \cite{Rosenow2019} give a consistent picture in which the statistical phase of anyonic quasiparticles contributes to the interference phase. Our data point to the impact of braiding both in the regime where the chemical potential is near the center of the energy gap, where the density of states is small and individual quasiparticle transitions can be resolved, and in the regimes above and below the gap where the density of states is high and a continuum of quasiparticle and quasihole states contribute to the phase.

The arguments in \cite{Rosenow2019} of a transition from a regime of constant $\nu$ to regimes of constant $n$ when moving away from the center of the state also apply to integer quantum Hall states. In Supp. Fig. 6 we show measurements of interference as a function of $B$ and $V_g$ across the integer state $\nu = 1$; in contrast to the fractional $\nu = 1/3$ case, the device exhibits no change in behavior and displays negatively-sloped Aharonov-Bohm interference at the high and low field extremes of the plateau. This is consistent with the fact that the charge carriers and excited states are electrons which obey fermionic statistics, making their braiding unobservable; $\theta_{fermion}=2\pi$.

\section{Temperature dependence}
\begin{figure}[t]
\def\ffile{MainTextFigures/TempDependence2.pdf}
\centering
\includegraphics[width=\linewidth]{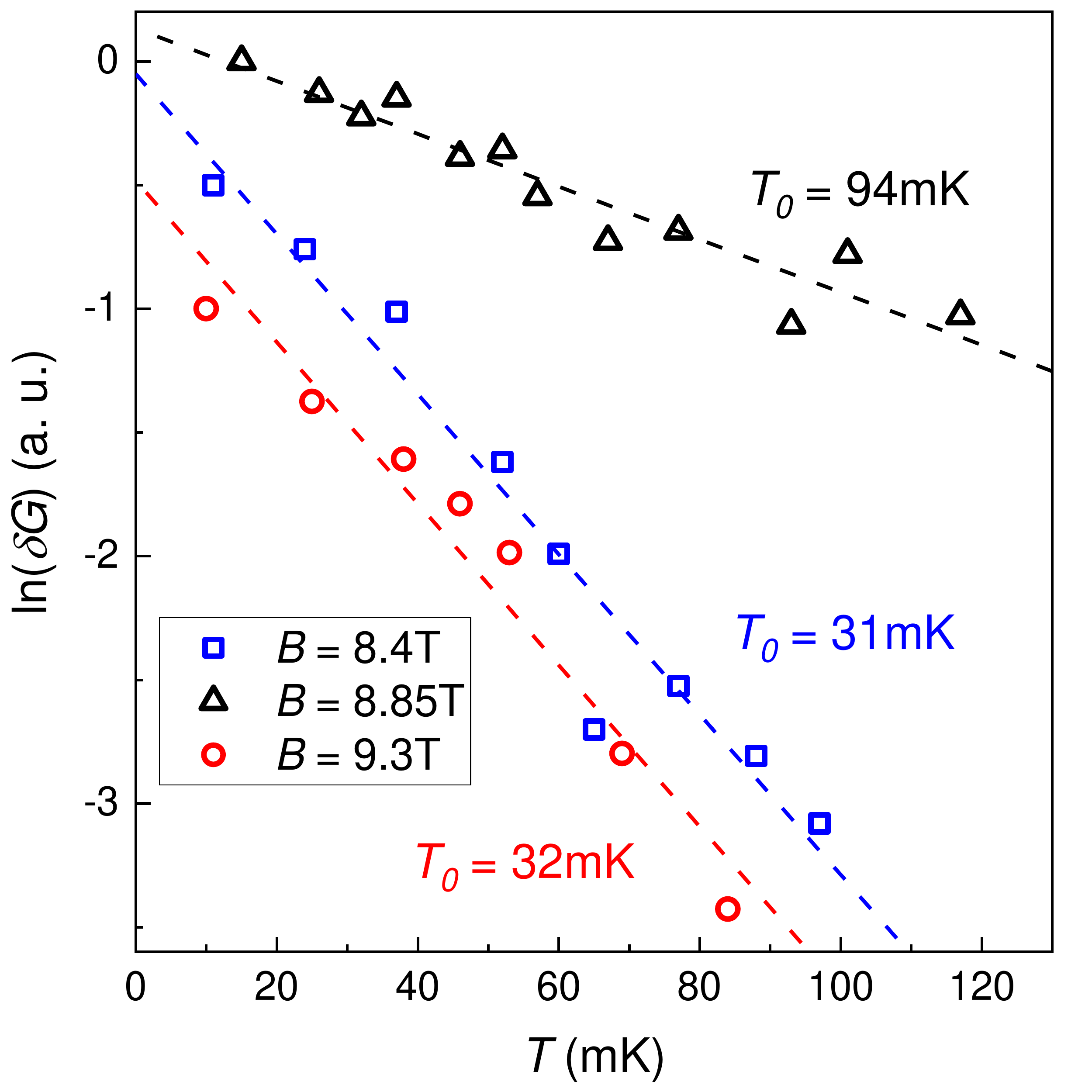}
\caption{\label{TempDependence} \textbf{Dependence of oscillation amplitude on temperature.} The natural log of the oscillation amplitude $\delta G$ at 8.4T, 8.85T, and 9.3T is plotted versus temperature. Data points are normalized to the amplitude at the lowest temperature and offset for clarity. The oscillation amplitudes show an approximately exponential decay with increasing temperature. Dashed lines indicate linear fits from which the temperature decay scale $T_0$ is extracted at each magnetic field. $T_0$ is much larger in the central region than in the low and high field regions, suggesting that there is an additional dephasing mechanism in these regions. This may be explained by topological dephasing due to thermal smearing of the quasiparticle number. The QPCs are tuned to approximately 90\% transmission at each temperature to maintain constant backscattering.}
\end{figure}

An additional observation is that the oscillation amplitudes decay with temperature much more sharply in the high-field and low-field regions than in the central region. We measured the amplitude of the oscillations in each region versus temperature; the oscillations decay approximately exponentially with $T$ as temperature increases, and we can characterize each region by the temperature decay scale $T_0$ assuming that the oscillation amplitude varies as $e^{\frac{-T}{T_0}}$ \cite{Chamon1997, Bishara2008,  Bishara2009, Hu2009}. We extract $T_0$ through a linear fit of the natural log of the oscillation amplitude as a function of temperature; this data is shown in Fig. \ref{TempDependence}. For the low-field region at 8.4T (blue) $T_0=31$mK, for the central region at 8.85T (black) $T_0=94$mK, and for the high-field region at 9.3T (red) $T_0=32$mK. Differential conductance measurements to extract the edge-state velocity were performed \cite{Chamon1997, McClure2009,  Heiblum2016}, and indicate that the edge velocity does not vary significantly between the different regions (see Supp. Discussion 3 and Supp. Fig. 7). Evidently, the observed suppression of $T_0$ by nearly a factor of 3 in the high and low field regions cannot be explained by lower edge velocity. Based on the measured edge velocities we calculate predicted temperature decay scales $T_0$ of 76mK at 8.4T, 89mK at 8.85T, and 85mK at 9.3T (see Supp. Discussion 2). In the central region the predicted value of $T_0$ is close to the experimentally observed value, indicating that in this region of constant $\nu$ and a small number of quasiparticles, dephasing can primarily be attributed to thermal smearing of the edge state based on dwell time in the interferometer. In the high and low field regions, however, the experimental values are much smaller than the predicted values, suggesting that there must be an additional source of dephasing in these regions. The fact that this increased dephasing occurs in the regions where a large number of quasiparticles and quasiholes populate the interferometer, but not in the central region, suggests that it may be explained by the topological dephasing proposed in \cite{Gefen2015}, in which thermal fluctuations in localized quasiparticle number reduce interference visibility in Fabry-Perot interferometers. This affirms the expectation that the regimes of constant density correspond to high quasiparticle DOS \cite{Rosenow2019}. This dephasing is a remarkable example of the non-local influence of anyonic statistics: despite the fact that the edge quasiparticles are well separated by many magnetic lengths from quasiparticles inside the bulk of the interferometer such that there is minimal direct interaction, thermal fluctuations in $N_{qp}$ nevertheless lead to rapid thermal dephasing of the interference signal.

The behavior of the device described here in the main text was reproduced in a second device, including the change in interference behavior from negatively-sloped Aharonov-Bohm interference to flat lines of constant phase, the suppression of $T_0$ outside the central region, and the observation of discrete phase jumps consistent with the predicted anyonic phase at $\nu=1/3$; see Supp. Fig. 8. The possible influence of residual bulk-edge interaction is discussed in Supp. Discussion 4.

\section{Conclusions}
We have measured conductance oscillations in a Fabry-Perot interferometer across a wide range of magnetic field at the $\nu = 1/3$ quantum Hall state. Near the center of the state, we observe discrete jumps in the interference phase consistent with the anyonic braiding statistics of localized quasiparticles, and we obtain $\theta _{anyon}={2\pi}\times(0.31 \pm 0.04)$, which agrees with the theoretically predicted value of $\theta_{anyon}=\frac{2\pi}{3}$. When the magnetic field is moved away from the center, we observe a change in interference behavior from predominantly negatively sloped lines of constant phase to a phase that is nearly independent of $B$. This observation suggests that the 2DES transitions from a regime of constant filling factor at the center to regimes  of constant density leading to a thermally smeared population of quasiparticles (at low field) and quasiholes (at high field), as predicted in a recent theoretical work \cite{Rosenow2019}. Extraction of the lever arms which parameterize the effects of gate voltage on charge in the device increases our confidence in this analysis. In the low and high field regimes we observe a dramatic increase in thermal dephasing evidenced by the suppression the temperature decay scale $T_0$, which indicates that despite their large spatial separation from the interfering edge state, localized quasiparticles have a profound impact on interference behavior through their braiding statistics. Taken together, our experimental observations are consistent with interference of anyonic quasiparticles for which braiding statistics contribute to the interference phase. 

\section{Methods} 
The device used was fabricated using the following steps: (1) optical lithography and wet etching to define the mesa; (2) deposition and annealing of Ni/Au/Ge Ohmic contacts; (3) electron beam lithography and electron beam evaporation (5nm Ti/10nm Au) to define the interferometer gates; (4) optical lithography and electron beam evaporation (20nm Ti/150nm Au) to define bondpads and surface gates around the Ohmic contacts; (5) mechanical polishing to thin the GaAs substrate; (6) optical lithography and electron beam evaporation (100nm Ti/150nm Au) to define the backgates used to deplete the bottom screening well around the Ohmic contacts so that only the primary quantum well is probed.

Standard low-frequency ($f$ = 13Hz) 4-terminal and 2-terminal lock-in amplifier techniques were used to probe the diagonal resistance and conductance across the device. Typically a 50pA excitation current was used for measurements. A +600mV bias was applied to the QPC and side gates while the device was cooled from room temperature; this bias-cool technique results in an approximately 600mV built-in bias on these gates, which was found to improve device stability.

\section{acknowledgements}
 This work is supported by the U.S. Department of Energy, Office of Science, Office of Basic Energy Sciences, under award number DE-SC0020138. G. C. Gardner acknowledges support from Microsoft Quantum.  The authors thank Bernd Rosenow for valuable comments on an early version of this manuscript.

\end{document}


\title{Supplementary Information for ``Direct observation of anyonic braiding statistics at the $\nu$=1/3 fractional quantum Hall state''}
\renewcommand{\figurename}{SUPP. FIG.}

\maketitle

\begin{figure*}[t]
\def\ffile{SupplementaryFigures/Structure.pdf}
\centering
\includegraphics[width=0.5\linewidth]{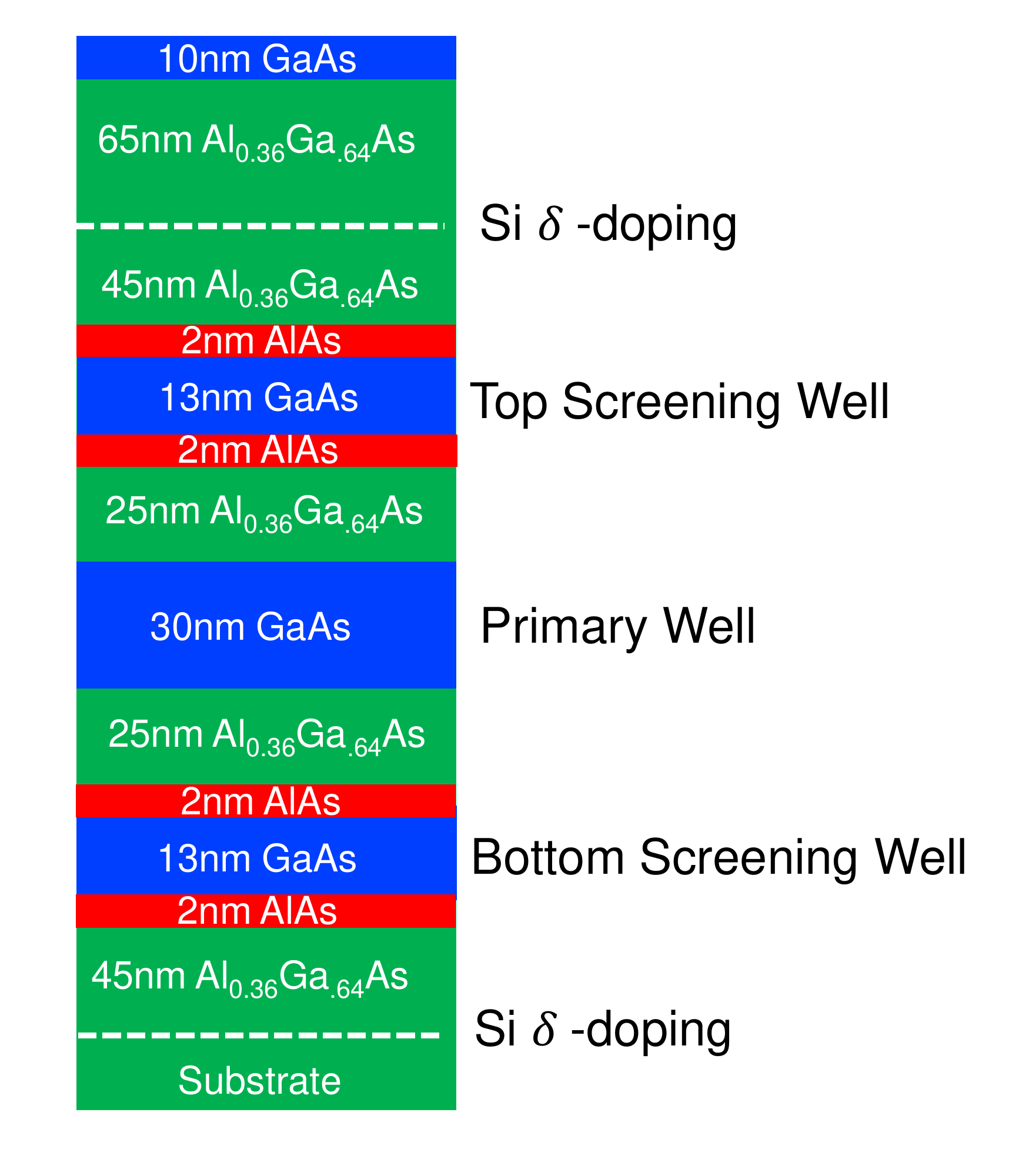}
\caption{\label{Structure} \textbf{Layer stack of the GaAs/AlGaAs heterostructure used for the experiments.} This structure utilizes three GaAs quantum wells: a primary 30nm well flanked by two 13nm screening wells to reduce the bulk-edge interaction in the interferometer. There are 25nm AlGaAs barriers between the main well and screening wells, and the total center-to-center setback of the screening wells from the main well is 48nm.}
\end{figure*}

\begin{figure*}[t]
\def\ffile{SupplementaryFigures/Repeatability2.pdf}
\centering
\includegraphics[width=0.7\linewidth]{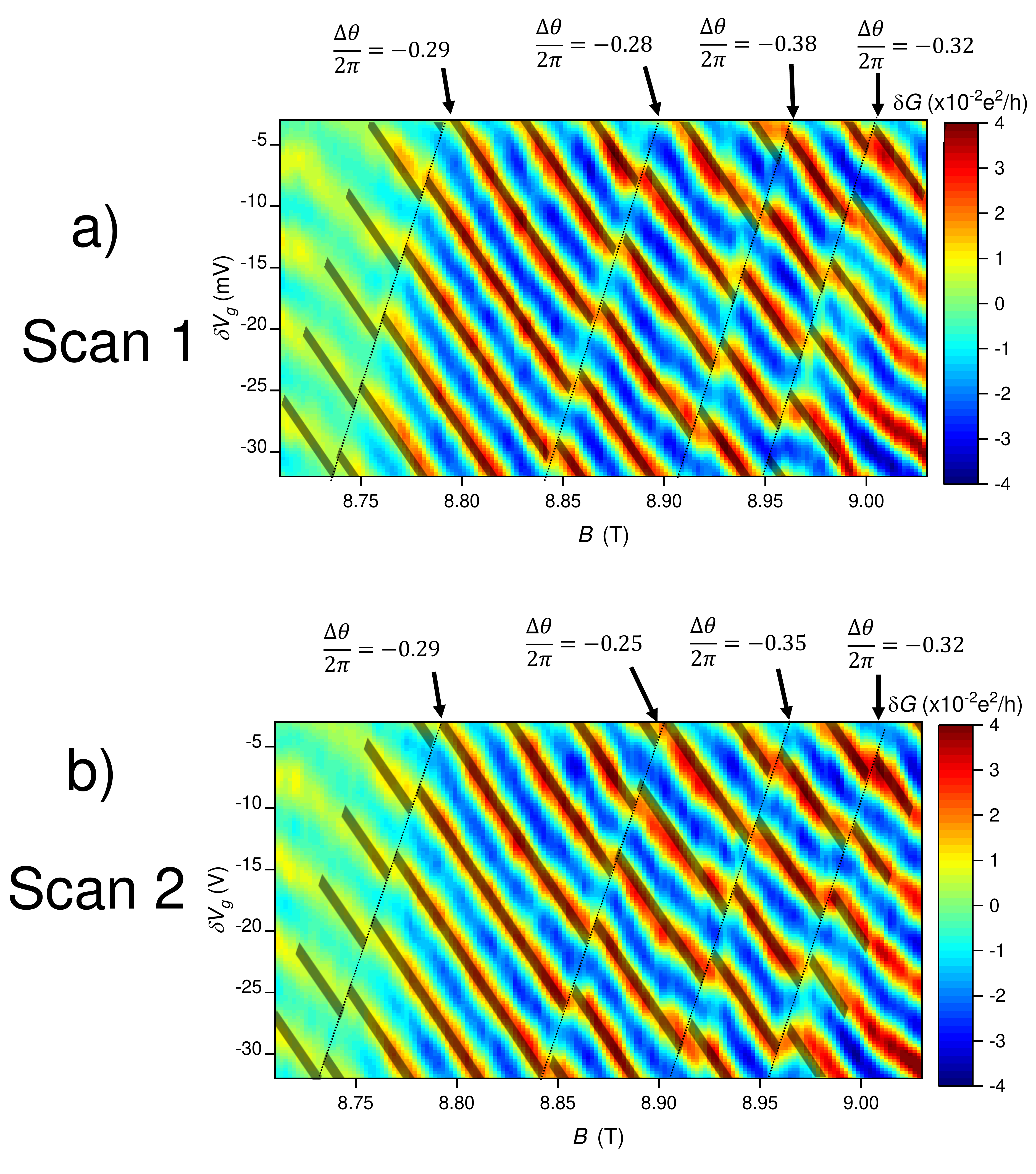}
\caption{\label{Repeat} \textbf{Repeatability of discrete phase jumps.} a)First scan measurement of conductance versus $B$ and $\delta V_g$. This is the same data in Fig. 2 of the main text. b) Second scan across the same range of magnetic field using the same QPC gate voltages. As can be seen from the data, the same pattern of discrete jumps appear in the second scan. The second scan was taken approximately one hour after the first scan. Values of $\frac{\Delta \theta}{2\pi}$ extracted from least squares fits are shown for both scans, and show similar values for each phase jump in both scans.}
\end{figure*}

\begin{figure*}[t]
\def\ffile{SupplementaryFigures/GapMeasurement.pdf}
\centering
\includegraphics[width=0.7\linewidth]{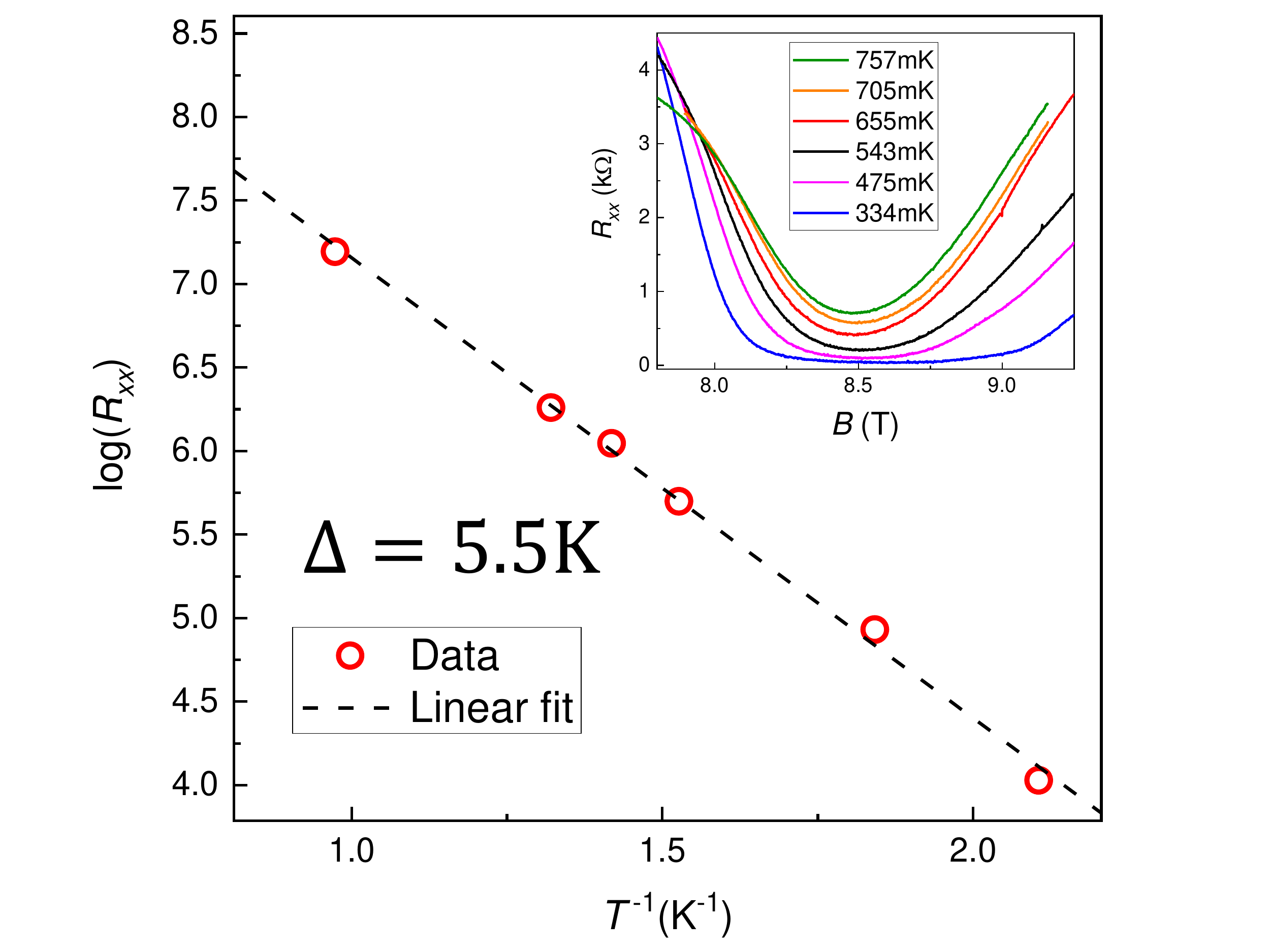}
\caption{\label{Gap} \textbf{Measurement of the energy gap for the $\nu = 1/3$ fractional quantum Hall state.} The inset shows longitudinal resistance $R_{xx}$ measured in a bulk region away from the interferometer at different temperatures. A linear fit of the data to the form $R_{xx} = R_0 e^{\frac{-\Delta}{2kT}}$ yields a gap of $\Delta =$5.5K. This is consistent with values measured in previous experiments at similar magnetic field \cite{Stormer1993}.}
\end{figure*}

\clearpage

\section{Supplementary Discussion 1: Analysis of periods and lever arms}
We discuss the change in interference behavior from the central region, where the device exhibits primarily Aharonov-Bohm interference behavior with a few phase jumps, to the high and low field regions, where the lines of constant phase flatten out. Interestingly, despite the lines of constant phase remaining continuous, the side gate oscillation period becomes smaller in the high and low field regions relative to the central region, with periods of 5.8mV at 8.4T, 8.5mV at 8.85T, and 5.4mV at 9.3T. In Supp. Fig. \ref{PeriodAnalysis}a line cuts of conductance versus gate voltage are shown at these magnetic fields illustrating the change in period. On the other hand, the model of \cite{Rosenow2019} suggests that the side gate oscillation period will be the same in the central region as in the upper and lower field regions, because the authors assume that the side gate couples only to the edge of the interferometer; under this assumption the side gate voltage will only affect the Aharonov-Bohm phase, making the variation of $\theta$ with $V_g$ the same in each region. However, in a real device the gates do not affect just area; they also have some effect on the charge in the bulk of the interferometer. In the high and low field regions this will lead to an additional change in phase with gate voltage due to changes in localized quasiparticle number. To analyze the effect this will have on the side gate oscillation period, in Supp. Eqn. \ref{PhaseLeverArm} we take the derivative of $\theta$ (from Eqn. 1 in the main text) with respect to side gate voltage: 

\begin{figure*}[t]
\def\ffile{SupplementaryFigures/Period_Analysis2.pdf}
\centering
\includegraphics[width=\linewidth]{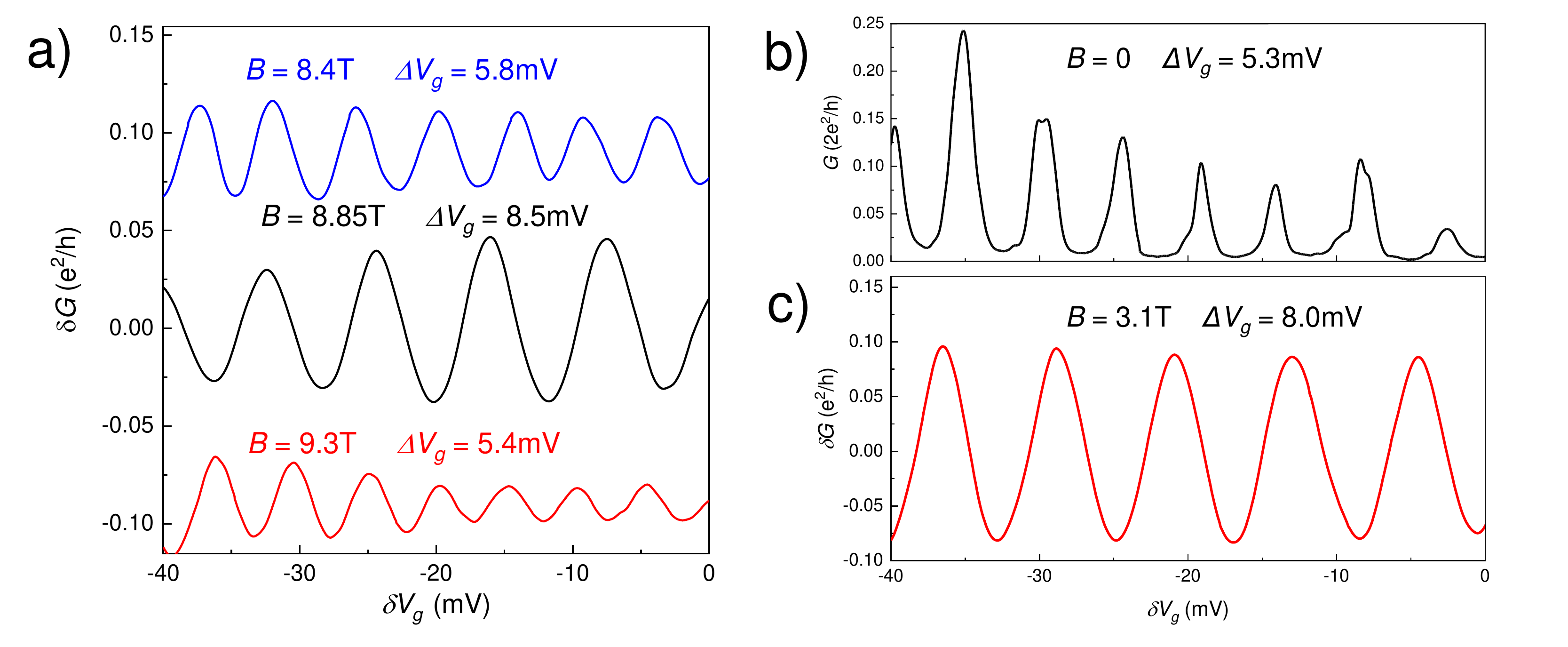}
\caption{\label{PeriodAnalysis} \textbf{Conductance oscillations at different magnetic fields.} a) Conductance oscillations $\delta G$ versus side gate voltage $\delta V_{g}$ in the low-field region at $B = 8.4$T (blue), in the central region at $B = 8.85$T (black), and in the high-field region at $B = 9.3$T (red). The side gate oscillation period $\Delta V_{side gates}$ is significantly smaller in the low field and high field regions than in the central region, with $\Delta V_{g}=5.8$mV at 8.4T, $\Delta V_{g} = 8.5$mV at 8.85T, and $\Delta V_{g}=5.4$mV at 9.3T. The QPCs are tuned to approximately 90\% transmission. b) Conductance $G$ versus side gate voltage at zero magnetic field with the device operated in the Coulomb blockade regime. Unlike other data presented in this work, the oscillations shown here are due to resonant tunneling of electrons rather than interference, and the QPCs are tuned weak tunneling, $G<<\frac{e^2}{h}$. The Coulomb blockade oscillations have a period of 5.3mV, which is used to obtain the total lever arm $\alpha_{total}$ of the gates to the interferometer. c) Aharonov-Bohm interference oscillations at $\nu = 1$. The oscillations period of 8.0mV is used to obtain the lever arm $\alpha _{edge}$ of the gates to the edge.}
\end{figure*}

\begin{equation} \label{PhaseLeverArm}
\frac{\partial \theta}{\partial V_{g}} = 2\pi \frac{e^*}{e} \frac{B}{\Phi _0}\frac{\partial A_I}{\partial V_{g}}+\theta _{anyon}\frac{\partial \langle N_L \rangle}{\partial V_{g}} 
\end{equation}

Here with $\langle N_L \rangle$ we take the thermally averaged number of quasiparticles to account for the fact that in the high and low field regimes significant thermal smearing is expected, and $\langle N_L \rangle$ will not necesarily be an integer \cite{Rosenow2019}. 

In order to determine whether change in localized quasiparticle number with gate voltage can explain the observed change in period, we determine the parameter $\alpha _{bulk}\equiv \frac{\partial q_{bulk}}{\partial V_g}$ which paramaterizes how the bulk charge inside the interferometer $q_{bulk}$ changes with $V_g$. To determine $\alpha _{bulk}$ we have operated the device at zero magnetic field in the Coulomb blockade regime \cite{Beenakker1991} with the QPCs tuned to weak tunneling; in this regime there is one conductance peak each time the number of electrons in the device changes by one. Inverting the Coulomb blockade oscillation period gives the total lever arm coupling the side gates to the interferometer, $\alpha _{total}=\frac{1}{\Delta V_{CB}}=\frac{\partial q_{total}}{\partial V_g}$. Zero-field Coulomb blockade oscillations are plotted in \ref{PeriodAnalysis}b; the 5.4mV period yields $\alpha _{total}=$0.19mV$^{-1}$. However, $q_{total}$ is a combination of charge at the edge and charge in the bulk, $q_{total}=q_{edge}+q_{bulk}$, so to determine $\alpha _{bulk}$ we must also determine $\alpha_{edge}\equiv \frac{\partial q_{edge}}{\partial V_g}$. To extract $\alpha_{edge}$ we operate the device as an Aharonov-Bohm interferometer at the integer quantum Hall state $\nu = 1$; in this regime the interference phase and thus the oscillation period depends only on change in interference area and not on changes in charges localized in the bulk \cite{Halperin2011}. Since for an integer state each oscillation period corresponds to changing the enclosed flux by one, $\alpha_{edge}=n\frac{\partial A_I}{\partial V_g}=\frac{n \Phi _0}{B} \frac{1}{\Delta V_{\nu = 1}}$ with $n$ the electron density (we assume that the electrostatics which determine the coupling of the gate to the edge and to the bulk do not change significantly with magnetic field). AB interference oscillations for the integer state $\nu = 1$ at B = 3.1T are show in Supp. Fig. \ref{PeriodAnalysis}c; the period of 8.0 mV gives $\frac{\partial A}{\partial V_g}=$ 0.167$\mu$m$^2$V$^{-1}$ and $\alpha_{edge}=0.12$mV$^{-1}$. Finally, we calculate $\alpha _{bulk} = \alpha _{total} - \alpha _{edge} = $0.19mV$^{-1}-$0.12mV$^{-1}=$0.07mV$^{-1}$. The fact that $\alpha _{edge}$ is significantly larger than $\alpha _{bulk}$ indicates that, as expected, the primary action of the side gates is to change the area of the interferometer, and the change in bulk charge is comparatively small.

We calculate the expected periods in each regime at $\nu=1/3$ by $\Delta V_g = 2\pi (\frac{\partial \theta}{\partial V_{g}})^{-1}$. In the central region quasiparticles are unlikely to be created because of the large energy gap, so only the first term on the right-hand side of Supp. Eqn. \ref{PhaseLeverArm} contributes, whereas in the low-field/high-field regions quasiparticles/quasiholes will be created and contribute to the phase, so both terms will contribute. For the central region then the predicted period is $\Delta V_g = \frac{\Phi_0}{B}\frac{e}{e^*}(\frac{\partial A_I}{\partial V_g})^{-1}\approx 8.4$mV, in good agreement with the measured value of 8.5mV. This agreement with the model suggests that interference in the central region can indeed be understood as the Aharononv-Bohm effect at constant $\nu$ of $e/3$ quasiparticles, consistent with theoretical predictions of fractional charge \cite{Laughlin1983} as well as previous experiments in interferometry \cite{Nakamura2019, Heiblum2010} and other experimental observations of fractional charge \cite{Goldman1995Science, Heiblum1997, Saminadayar1997}. At 8.4T and 9.3T, taking into account the creation of quasiparticles, we calculate $\Delta V_g = \frac{1}{ \frac{B}{\Phi_0}\frac{e^*}{e}\frac{\partial A_I}{\partial V_g}+\frac{\theta _{anyon}}{2\pi} \frac{\partial N_L}{\partial V_g}}=\frac{1}{ \frac{B}{3 \Phi_0} \frac{\partial A_I}{\partial V_g}+\alpha_{bulk}}$; here we have used $\frac{e^*}{e}=\frac{1}{3}$, $\theta _{anyon} = \frac{2\pi}{3}$, and $\frac{\partial N_L}{\partial V_g}=\frac{e}{e^*} \alpha_{bulk}$. This equation yields a predicted $\delta V_g$ of $\approx 5.5$mV at 8.4T and $\approx 5.1$mV at 9.3T, in good agreement with the experimental values of 5.8mV and 5.4mV. This agreement between predicted and observed oscillation periods in each region is strong support for the picture in \cite{Rosenow2019} of a region of constant $n$ and a quasiparticle population at low field, a region of constant $\nu$ near the center of the state, and a region of constant $n$ and a quasihole population at high field.

Additionally, these extracted lever arms can be used analyze quantitatively the slope of the quasiparticle transition lines in Fig. 2 of the main text. These transitions will occur when it becomes energetically favorable for a quasiparticle to be created at a certain place in the device, therefore the transition lines will correspond to lines of constant electrostatic energy associated with charge on the device. This electrostatic energy comes from accumulating charge on the device due to changes in the condensate charge density with magnetic field, which may be compensated for by the creation of quasiparticles. The charge on the 2DES is thus a combination of the condensate charge density and the charge associated with each localized quasiparticle:

\begin{equation} \label{TotalCharge}
q_{2DES} = \frac{e \nu A_I B}{\Phi_0}+e^* N_{qp}
\end{equation}

Furthermore, we must consider the net charge, $q_{net}$, the difference between the charge in the 2DES (from Supp. Eqn. \ref{TotalCharge}) and the background charge:

\begin{equation} \label{NetCharge}
q_{net} = q_{2DES}-q_{back}= \frac{e \nu AB}{\Phi_0}+e^* N_{qp}-q_{donor}-e\alpha_{bulk}\delta V_g
\end{equation}

Here the background charge is a combination of the charge from the donors, $q_{donor}$, and the effect of the gate voltage. We follow \cite{Halperin2011} in treating the gates as creating some effective additional background charge $e\alpha_{bulk}\delta V_g$. Since the changes in localized quasiparticle number occur when the electrostatic energy cost exceeds the energy cost to create a quasiparticle, Supp. Eqn. \ref{NetCharge} implies that the localized quasiparticle transitions will occur across lines with a slope $\frac{dV_g}{dB}=\frac{\nu A}{\Phi_0 \alpha _{bulk}}$. Using $\nu  = 1/3$, $\alpha_{bulk} = 0.07$mV$^{-1}$ (discussed above), and area extracted from the AB oscillations at $\nu = 1$ $A=\frac{\Phi_0}{\Delta B_{\nu = 1}}\approx0.38\mu m^2$ (Supp. Fig. \ref{v=1}b), we obtain $\frac{dV_g}{dB} \approx 0.44$mV/mT. Experimentally, the observed phase jumps occur with a slope of approximately 0.5 mV/mT, in good agreement with the predicted value. This is strong evidence that the discrete phase jumps do indeed correspond to changes in the number of localized quasiparticles inside the interferometer.

\clearpage

\section{Supplementary Discussion 2: Simulations}

\begin{figure*}[t]
\def\ffile{SupplementaryFigures/Simulations2.pdf}
\centering
\includegraphics[width=0.8\linewidth]{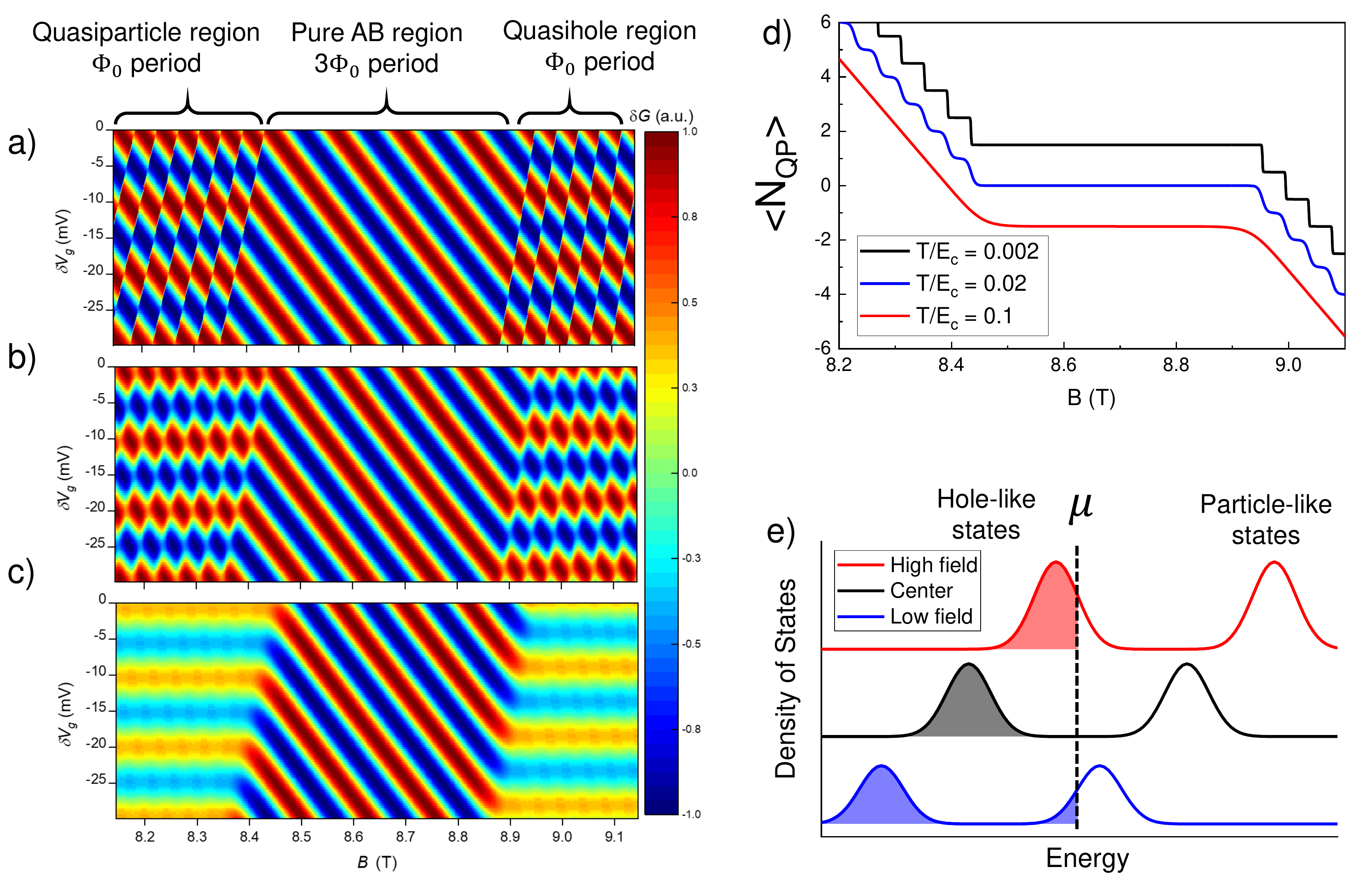}
\caption{\label{Simulations} \textbf{Simulations of interferometer behavior at $\nu = 1/3$.} Conductance values are computed as a function of magnetic field $B$ and  side gate voltages $V_{g}$, taking into account both the Aharonov-Bohm phase and the contribution $\theta _{anyon}$ from braiding around localized quasiparticles inside the bulk of the interferometer. Simulations are performed at different ratios of the temperature $k_BT$ to the interferometer charging energy $E_c=\frac{e^2}{2C}$ a) 0.002 b) 0.02 and c) 0.1. d) Plot of the thermal expectation value of the number of localized quasiparticles inside the interferometer for different ratios of $k_BT/E_c $; in this context a negative quasiparticle number indicates a population of quasiholes. In each case in the middle of the state there are no quasiparticles, resulting in conventional Aharonov-Bohm interference with $3\Phi _0$ period, while at higher fields quasiholes form and at lower fields quasiparticles form, resulting in phase slips with $\Phi_0$ period. As temperature is elevated, the quasiparticle number is thermally smeared, making the $\Phi_0$ period phase slips unobservable and reducing the amplitude of the oscillations that occur as a function of $V_{g}$. e) Qualitative plot of the density of states versus energy. }
\end{figure*}

To further validate application of the model of \cite{Rosenow2019} to our experimental results, we have performed simulations of interferometer behavior to model the conductance versus gate voltage $V_g$ and magnetic field $B$. The starting point for this is the equation for the interference phase difference which determines conductance oscillations:

\begin{equation} \label{FractAB}
\theta = 2\pi \frac{e^*}{e} \frac{A_I  B}{\Phi _0}+N_{qp} \theta _{anyon}
\end{equation}

The conductance for an interferometer varies as $\delta G= a_0 cos(\theta)$, with $a_0$ an amplitude that depends on the backscattering of the QPCs. In our system thermal fluctuations in $N_{qp}$ may be important, so we need to calculate a thermal average (because we use low-frequency measurement techniques, this thermal average should correspond to the experimentally measured conductance). Following \cite{Halperin2011}, we compute the thermal expectation value $\langle \delta G \rangle$:

\begin{equation} \label{ThermalAverage}e
\displaystyle \langle \delta G \rangle = \frac{1}{Z} \sum_{N_{qp}=-\infty}^{+\infty}e^{\frac{-E(N_{qp})}{k_BT}} \cos{(2\pi \frac{e^*}{e} \frac{A_I  B}{\Phi _0}+N_{qp} \theta _{anyon})}
\end{equation}

\begin{equation} \label{Partition}
Z = \sum_{N_{qp}=-\infty}^{+\infty}e^{\frac{-E(N_{qp})}{k_BT}} 
\end{equation}

Here $E$ is the energy of the device as a function of $N_{qp}$. As in the main text, a negative quasiparticle number here corresponds to a population of quasiholes. We define an energy similar to that in \cite{Rosenow2019}, but for the interferometer rather than the bulk energy per unit area:

\begin{equation} \label{Energy}
E = E_0 + \frac{e^2}{2C}\delta q_{net}^2+\Delta_{qp}  |N_{qp}| 
\end{equation}

$E_0$ is an offset accounting for the energy of the condensate which does not depend on the number of quasiparticles, so it is left out of the simulation. The absolute value of $N_{qp}$ is multiplied by $\Delta_{qp}$ to account for the fact that quasiholes (corresponding to negative $N_{qp}$) also cost energy $\Delta _{qp}$, and for simplicity we set the energy associated with creating a quasiparticle $\Delta_{qp}$ to be half of the full gap, $\Delta_{qp}=\frac{\Delta}{2}$. This is a simplification which assumes that the energy cost for creating quasiparticles is the same as for quasiholes, although numerical results have indicated that quasiparticles have a higher energy cost than quasiholes \cite{MacDonald1986}. Nevertheless, the width of the constant $\nu$ region over which no quasiparticles or quasiholes are created is determined by the full gap which is a sum of the quasiparticle and quasihole gaps, $\Delta=\Delta_{qp}+\Delta_{qh}$, and this full gap is the value measured in transport. So, asymmetry in the energies of quasiparticles and quasiholes do not affect quantitative comparison between the theory of \cite{Rosenow2019} and our experimental results.

As discussed in the Supp. Discussion 1, $q_{net}$ is the difference between the charge in the 2DES and the compensating background charge (including that created by the side gate), and is written in Supp. Eqn. \ref{NetCharge}. The term $\frac{e^2}{2C}\delta q_{net}^2$ in Supp. Eqn. \ref{Energy} gives the energy cost associated with building up excess charge  on the device. In the presence of screening layers separated from the main well by distance $d$, we estimate the characteristic capacitance to be $C=\frac{2\epsilon A_I}{d}$.

Simulations of conductance versus magnetic field and side gate voltage are performed by numerically evaluating Supp. Eqn. \ref{ThermalAverage} and Supp. Eqn. \ref{Partition} at each value of $B$ and $\delta V_g$. The area $A_I$ is computed as $A_I=A_0+\frac{\partial A}{\partial V_g}\delta V_g$. Rather than performing an infinite sum, we sum over $N_{qp}$ from  -20 to +20 to make the computation possible; this is justified because states with large numbers of quasiparticles are exponentially suppressed. Simulations are performed for the $\nu = 1/3$ state, so based on theoretical expectations we set $\theta_{anyon} = \frac{2\pi}{3}$ and $e^*=e/3$. The value of 5.5K for $\Delta$ extracted from the bulk transport gap measurement in Supp. Fig. \ref{Gap} was used, giving a value of 2.75K for $\Delta_{qp}$. A 2DES of $0.7\times 10^{11} cm^{-2}$ is assumed, which sets the background charge $q_{donor}$.  

In Supp. Fig. \ref{Simulations}a, b, and c we show the results of the simulations at different temperatures. The energy scale which primarily determines the thermal smearing effect of temperature is $\frac{e^2}{2C}$, so we set the ratio of $kT$ to $\frac{e^2}{2C}$ at 0.002, 0.02, and 0.1 in a, b, and c. In order to make the behavior in the simulations easier to see in the plots, these simulations are performed with a device with smaller area than the real device; we set $A_0$=0.1$\mu$m$^2$, whereas for the real device $A_0 \approx 0.38\mu$m$^2$ based on the Aharonov-Bohm periods. We first focus on the low-temperature simulation in a). Qualitative features of the simulation match the experiment: negatively-sloped Aharonov-Bohm oscillations of period $3\Phi_0$ occur near the center, and this behavior is confined to a 530mT region (consistent with the value calculated from the model in \cite{Rosenow2019} and close to the experimentally observed value of $\approx 450$mT). Above and below this region, the simulation in a) shows sharply defined discrete jumps in phase occurring with period $\Phi_0$, consistent with the findings of \cite{Rosenow2019}. As temperature is increased in b) and c), the transitions in these phase jumps become thermally smeared together, such that at the highest value of temperature simulated the transitions in phase are nearly completely smeared out, and the simulation shows nearly flat lines of constant conductance. This is in good qualitative agreement with the experimental results.

An additional subtle feature in the simulations is that the transition from the central Aharonov-Bohm region to quasiparticle and quasihole regions  occurs across a line with positive slope in the $V_g-B$ plane, due to the coupling of the side gate to the bulk included in the simulation, $\alpha_{bulk}$. This behavior is in fact observed in the experimental data in the transition to the high-field region. In the low-field transition this behavior is less clear because the transition appears to occur more smoothly rather than abruptly, but a positive slope is still observable. 

In Supp. Fig. \ref{Simulations}d, line cuts of the thermally averaged quasiparticle number $\langle N_{qp} \rangle$ are plotted versus $B$ from the simulations at each temperature.  At low temperatures this forms a staircase-like function with very sharp transitions when it becomes energetically favorable to change the number of localized quasiparticles, whereas at high temperatures these transitions become thermally smeared such that the evolution becomes quite smooth. While our simplified model and simulations likely do not capture all of the physics of the device, we believe that this picture of the average quasiparticle number becoming thermally smeared at high temperature smeared should hold. Additionally, it is possible that other mechanisms, such as charge noise, may result in smearing of $\langle N_{qp} \rangle$ on the measurement time scale.

\begin{figure*}[t]
\def\ffile{SupplementaryFigures/v=1.pdf}
\centering
\includegraphics[width=0.7\linewidth]{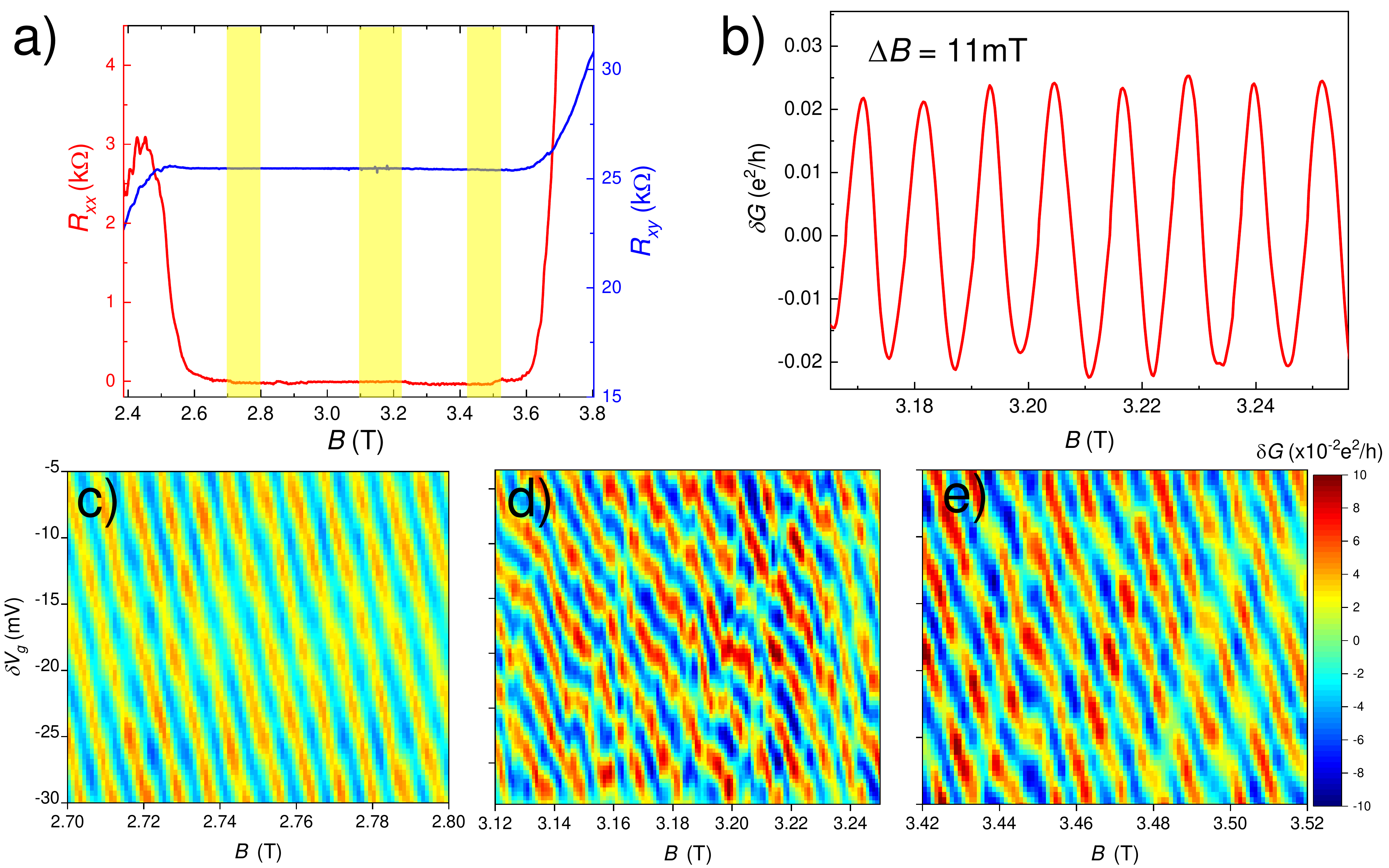}
\caption{\label{v=1} \textbf{Measurements of interference at $\nu = 1$.} a) Bulk quantum Hall transport showing the zero in $R_{xx}$ and plateau in $R_{xy}$ corresponding to the $\nu=1$ integer quantum Hall state. For this integer state, the bulk excitations and edge state current carrying particles are simply electrons, which obey fermionic statistics. b) Conductance oscillations versus magnetic field, showing an oscillation period  $\Delta B=$11mT.  From this period the effective area $A_I$ of the interferometer can be extracted: $A_I=\frac{\Phi_0}{\Delta B}$ In c), d), and e) we show conductance versus $B$ and $\delta V_g$ across the interferometer in the low field region of the plateau, near the center of the plateau, and on the high-field side of the plateau; the region on the plateau corresponding to each pajama plot is shown in a). In each of these regions the device exhibits negatively sloped Aharonov-Bohm oscillations. This contrasts with the data shown in the main text for the $\nu = 1/3$ state where lines of constant phase flatten out at high and low fields. This is consistent with the fact that electrons, which carry current and form localized states at $\nu = 1$, are fermions who obey trivial braiding statistics, $\theta _{fermion}=2\pi$, making braiding unobservable and leading to no change in interference behavior.}
\end{figure*}
\clearpage

\section{Supplementary Discussion 3: Velocity Measurements}

\begin{figure*}[t]
\def\ffile{SupplementaryFigures/Velocity_1-3.pdf}
\centering
\includegraphics[width=0.7\linewidth]{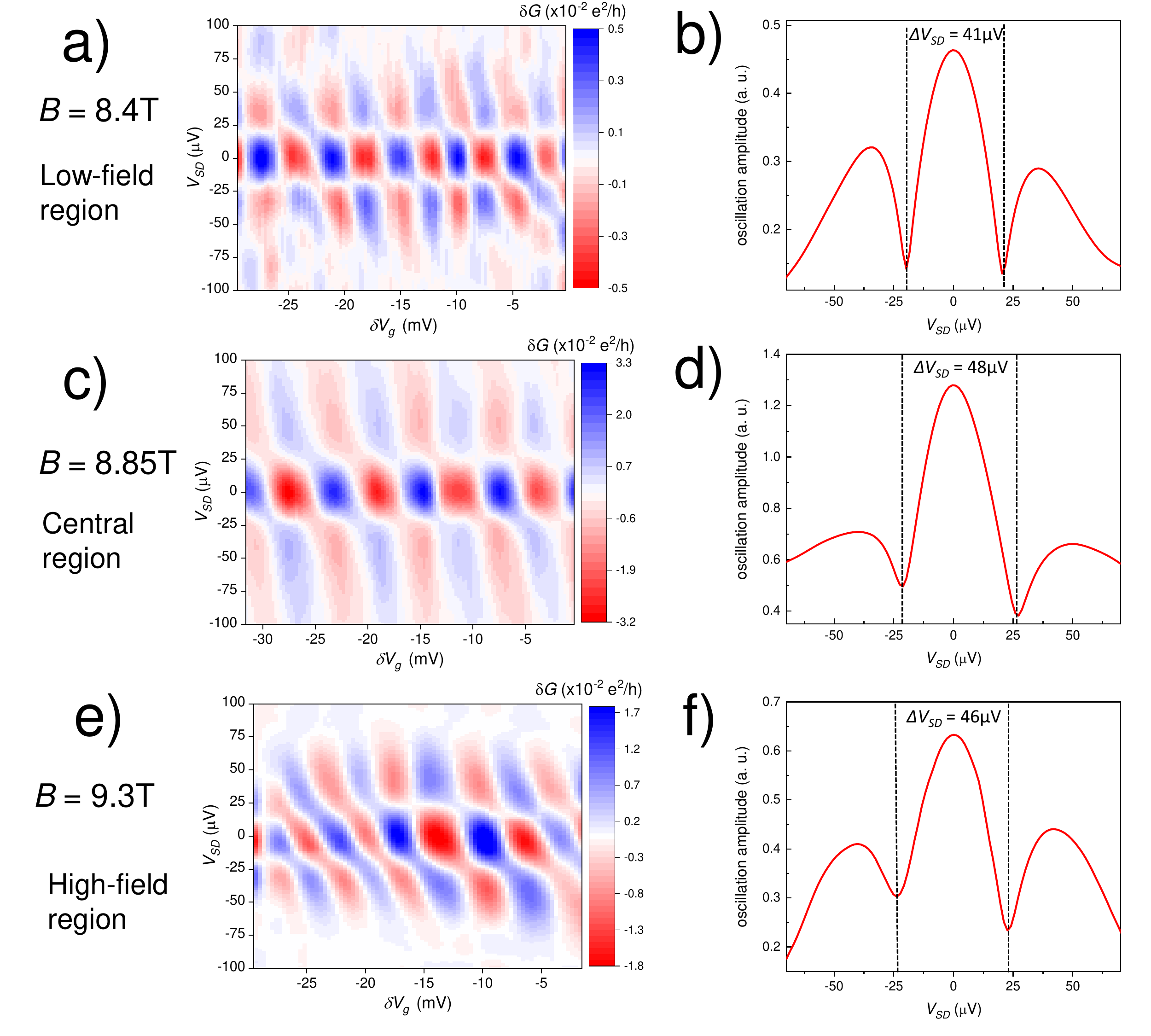}
\caption{\label{Velocity} \textbf{Differential conductance measurements at $\nu = 1/3$.} a) Differential conductance $\frac{\partial I}{\partial V_{sd}}$ as a function of side gate voltage $\delta V_g$ and source-drain bias $V_{sd}$ at $B = 8.4$T in the low-field region. b) Conductance oscillation amplitude from a FFT of the conductance versus side gate voltage data as a function of $V_{sd}$. The oscillation amplitude shows a node pattern as a function of $V_{sd}$ from which the edge velocity may be extracted, yielding $v_{edge}=8.3\times 10^3$m/s. c) Differential conductance and d) oscillation amplitude versus $V_{sd}$  at 8.85T giving $v_{edge}=9.7\times 10^3$m/s. e) Differential conductance and f) oscillation amplitude versus $V_{sd}$  at 9.3T giving $v_{edge}=9.3\times 10^3$m/s. Evidently, the edge velocity does not change significantly across the $\nu = 1/3$ quantum Hall plateau.}
\end{figure*}

In the main text we have discussed the observation that the temperature decay scale is much smaller in the high and low field regions than in the central region at $\nu = 1/3$, which suggests that topological dephasing due to thermal smearing of the localized quasiparticle number inside the interferometer may contribute in the high and low field regions. However, another possible explanation for the change in $T_0$ is that the edge velocity might be much larger at the center, and decrease in the high and low field regions; if this were the case, $T_0$ would decrease simply due to the additional thermal smearing of the edge state (that said, such a large and non-monotonic change in the velocity over such a relatively small range of $B$ would be rather surprising). In order to determine if the change in $T_0$ can be explained by changes in the edge velocity $v_{edge}$, we have performed differential conductance measurements in each region from which $v_{edge}$ can be extracted \cite{Chamon1997}; this has been performed previously for integer quantum Hall states in Aharonov-Bohm interferometers  \cite{McClure2009, Heiblum2016, Nakamura2019}. We consider first the case of an integer charge edge state. When a finite source drain voltage bias is applied, the energy of the injected edge electrons changes, and this leads to a  shift in the phase from the edge dispersion, $\delta \theta = \delta \epsilon \frac{\partial k}{\partial \epsilon}L = \frac{\delta \epsilon L}{\hbar v_{edge}} $. This leads to an additional interference pattern that occurs as a function of source-drain bias $V_{sd}$ which can be observed in the differential conductance measurement, resulting in a checkerboard pattern in the measured differential conductance as a function of $V_{sd}$ and $\delta V_g$. Thus nodes in the conductance oscillations which occur as a function of $\delta V_g$ occur at certain values of $V_{sd}$, and the spacing between nodes can be used to extract the edge velocity \cite{Chamon1997, Litvin2008, Heiblum2016, Nakamura2019}:

\begin{equation} \label{velocity}
v_{edge} = \frac{e^*L \Delta V_{sd}}{2\pi \hbar}
\end{equation}

Here $L$ is the perimeter of the interferometer, estimated based on the area extracted from Aharonov-Bohm interference measurents $L=4\sqrt{A_I}\approx  2.5 \mu$m. Differential conductance measurements are shown at $\nu = 1/3$ in Supp. Fig. \ref{Velocity}. Also plotted is the oscillation amplitude (extracted from a Fourier transform of the data) versus $V_{sd}$, which enables convenient extraction of $\Delta V_{sd}$ \cite{Heiblum2016}. This is shown for the low-field region in a) and b), for the central region in c) and d), and for the high-field region in e) and f). Using Supp. Eqn. \ref{velocity} yields edge velocities of $8.3\times 10^3$m/s in the low-field region at $B=8.4$T, $9.7\times 10^3$m/s in the central region at $B=8.85$T, and $9.3\times 10^3$m/s in the low-field region at $B=9.3$T. The fact that the velocity does not change significantly with magnetic field indicates that a change in edge velocity cannot account for the large change in temperature decay scale $T_0$ between regions. The expected $T_0$ can be calculated as  $T_0=\frac{h}{2\pi k_B \tau}\frac{1}{g}$  \cite{Chamon1997, Bishara2009} where $g$ is the scaling exponent of the edge state, $g=\frac{1}{3}$ for $\nu=1/3$ \cite{Chamon1997}, and $\tau=\frac{L}{v_{edge}}$ is the time for the edge state to traverse the interferometer. This yields predicted $T_0$ based on thermal smearing of the edge state of 76mK at $B=8.4$T, 89mK at 8.85T, and 85mK at 9.3T. The predicted value of 89mK in the central region at 8.85T is close to the experimentally observed $T_0$ of 94mK, indicating that the decay of amplitude in the region where the device is nearly free of quasiparticles can be attributed to thermal smearing of the edge. We have observed similar agreement between predicted and observed $T_0$ at the integer quantum Hall state $\nu = 1$ \cite{Nakamura2019}. However, the experimentally observed $T_0$ of 31mK in the low-field region and 32mK in the high field region at $\nu=1/3$ are much smaller than the values predicted for thermal smearing of the edge, indicating that another dephasing mechanism must be at play. This provides further support for the theory that topological dephasing due to thermally smearing of localized quasiparticles contributes to dephasing in these regions.

\begin{figure*}[t]
\def\ffile{SupplementaryFigures/DeviceB2.pdf}
\centering
\includegraphics[width=0.7\linewidth]{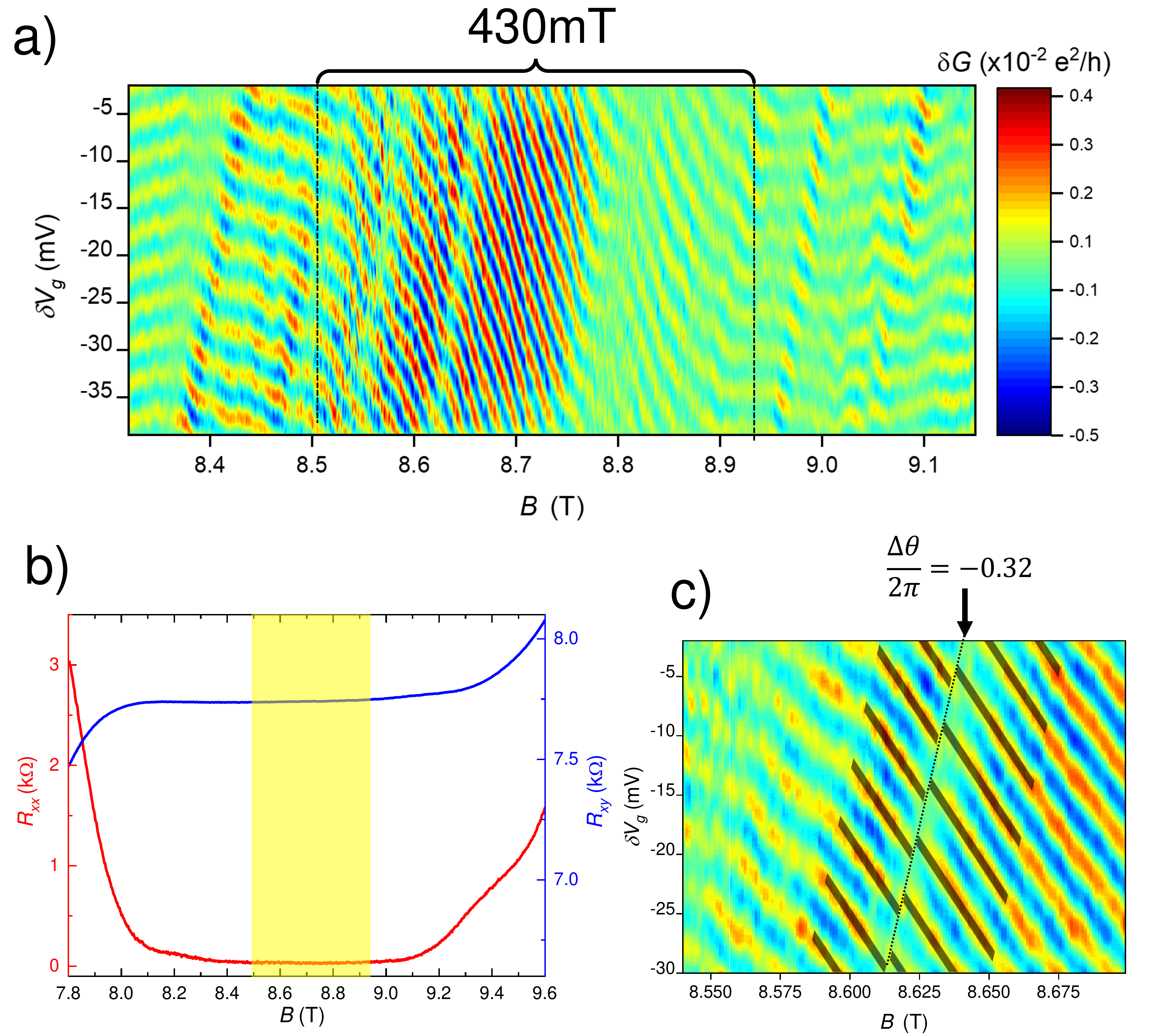}
\caption{\label{DeviceB} \textbf{Measurements of interference for a second device, taken from a different chip fabricated on the same wafer.} a) Conductance across the interferometer versus magnetic field $B$ and side gate voltage $\delta V_g$; $\delta V_g$ is relative to -1.0V. Behavior is similar to that observed in the device described in the main text: in a finite region with width $\approx 430$mT, the device exhibits negatively sloped Aharonov-Bohm oscillations, which flatten out at higher and lower magnetic fields, consistent with the creation of quasipaticles and quasiholes.  b) Bulk magnetotransport showing $R_{xx}$ (red) and $R_{xy}$ (blue) for device B. The region near the center of the $\nu = 1/3$ state where the negatively sloped Aharonov-Bohm oscillations occur is highlighted. c) zoomed-in view of a clear phase jump in the data (this jump is also visible in b), but the data in c) is a different scan intended to improve signal to noise). Least-squares fits of the conductance on either side of the phase jump yields an extracted phase jump $\frac{\Delta \theta}{2\pi}=-0.32$, yielding an anyonic phase $\theta_{anyon}=2\pi \times0.32$, consistent with theory.}
\end{figure*}

\clearpage

\section{Supplementary Discussion 4: Possible Bulk-edge  interaction effects}
We discuss another possible explanation for some of the experimental observations. It is worth mentioning that there is another mechanism which can cause discrete changes in phase in quantum Hall interferometers, even for integer quantum Hall states. In devices which are intermediate between the Aharonov-Bohm and Coulomb-dominated regimes, creation of a localized charge inside the interferometer causes the area of the interferometer to change due to finite bulk-edge coupling, resulting in a reduction in the Aharonov-Bohm phase and visible discrete changes in the interference phase. This mechanism does not depend on exotic braiding statistics \cite{Halperin2011}. However, increasing the magnetic field should tend to remove particle-like quasiparticles or create hole-like quasiparticles; in either case each excitation will lead to an an increase in phase when magnetic field is increased, because the decrease in $N_{qp}$ would be accompanied by compensatory increase in $A_I$. However, this is inconsistent with our observation of negative changes in phase across each discrete phase jump, and also inconsistent with the fact that these discrete phase jumps occur in the region where the Aharonov-Bohm phase shows clear negative slope, indicating minimal bulk-edge interaction. Nevertheless it is possible that some residual bulk-edge interaction may have a small effect on the observed phase jumps. In \cite{Halperin2011} it was found that the observed jumps in phase when changing quasiparticle number should be $\Delta \theta = \theta _{anyon} \times (1-\frac{K_{IL}}{K_I})$, where $\frac{K_{IL}}{K_I}$ is the ratio of the bulk-edge interaction strength $K_I$ to the characteristic energy cost for charging the edge $K_I$. Thus, residual bulk-edge interaction would result in a slightly smaller observed change in phase. This might account for the fact that the majority of the observed phase jumps are slightly smaller than $\frac{2\pi}{3}$.